\documentclass[12pt]{article}
\usepackage{graphicx}
\usepackage{mathrsfs}
\usepackage{amssymb}
\usepackage{amsmath}
\usepackage[margin=1cm]{caption}
\usepackage{verbatim}
\usepackage{slashed}
\newcommand{\be}{\begin{equation}}
\newcommand{\ee}{\end{equation}}
\newcommand{\bea}{\begin{eqnarray}}
\newcommand{\eea}{\end{eqnarray}}
\newcommand{\bmat}{\left(\begin{array}}
\newcommand{\emat}{\end{array}\right)}
\newcommand{\wino}{\widetilde{W}}
\newcommand{\bino}{\widetilde{B}}
\newcommand{\hdino}{\widetilde{\psi}^0_d}
\newcommand{\huino}{\widetilde{\psi}^0_u}
\newcommand{\sino}{\widetilde{\psi}^0_s}
\newcommand{\gev}{\text{GeV}}
\newcommand{\tev}{\text{TeV}}

\newcommand{\Mneut}{\mathcal{M_{\text{neut}}}}
\newcommand{\none}{\widetilde{\chi}^0_1}

\def\nulein{\nonumber \\}
\newcommand{\gsim}{\begin{array}{c}\sim\vspace{-21pt}\\> \end{array}}
\def\lsim{\:\raisebox{-0.5ex}{$\stackrel{\textstyle<}{\sim}$}\:}

\newcommand{\Mwino}{M_{\tilde{W}}}
\newcommand{\mue}{\mu_\text{eff}}

\newcommand{\Al}{A_\lambda}

\newcommand{\ra}{\rightarrow}

\newcommand{\Mchione}{M_{\tilde{\chi}^0_1}}
\newcommand{\Mchionepm}{M_{\tilde{\chi}^\pm_1}}

\newcommand{\MAonesq}{M^2_{A_D}}
\newcommand{\MAtwosq}{M^2_{A_S}}
\newcommand{\MAonesqbar}{\overline{M}_{A_D}^2}
\newcommand{\MAtwosqbar}{\overline{M}_{A_S}^2}
\newcommand{\Tneut}{T_{\chi}}
\newcommand{\muH}{\mu}

\title{
\vspace*{5mm} \Large\textbf{A 125 GeV Fat Higgs at large $\tan \beta$}
\vspace*{1.0cm}
\author{\textbf{Arjun Menon$^{a,b}$ and Nirmal Raj$^b$} \\
~\\
 \normalsize\emph{$^a$Institute of Mathematical Sciences, Taramani, Chennai, 600113, India}\\
 \normalsize\emph{$^b$Institute of Theoretical Science, University of Oregon, Eugene, OR 97405}
}
}

\date{\today}

\begin{document}
\setcounter{page}{0}
\maketitle
\abstract{In this article we study the viability of regions of large $\tan \beta$ within the framework of Fat Higgs/$\lambda$-SUSY Models.
We compute the one-loop effective potential to find the corrections to the Higgs boson mass due to the heavy non-standard Higgs bosons.
As the tree level contribution to the Higgs boson mass is suppressed at large $\tan \beta$, these one-loop corrections are crucial to raising the Higgs boson mass to the measured LHC value.
By raising the Higgsino and singlino mass parameters, typical electroweak precision constraints can also be avoided.
We illustrate these new regions of Fat Higgs/$\lambda$-SUSY parameter space by finding regions of large $\tan \beta$ that are consistent with all experimental constraints including direct dark matter detection experiments, relic density limits and the invisible decay width of the $Z$ boson.
We find that there exist regions around $\lambda = 1.25, \tan \beta = 50$ and a uniform psuedo-scalar $4\mbox{ TeV}\lsim M_A \lsim 8$~TeV which are consistent will all present phenomenological constraints. In this region the dark matter relic abundance and direct detection limits are satisfied by a lightest neutralino that is mostly bino or singlino.
As an interesting aside we also find a region of low $\tan \beta$ and small singlino mass parameter where a well-tempered neutralino avoids all cosmological and direct detection constraints.}
\thispagestyle{empty}
\newpage
\setcounter{page}{1}

\section{Introduction}

Weak scale supersymmetry (SUSY) remains a popular and elegant solution to the hierarchy problem of the Standard Model (SM)~\cite{Nilles:1983ge}.
It provides a natural means to stabilize the electroweak scale against large quadratic corrections from higher scales.
The fact that the Higgs boson was discovered to be light~\cite{Aad:2012tfa,Chatrchyan:2012ufa}, in accordance with SUSY's predictions, encourages us to continue our search for signals of SUSY at the TeV scale.

In the Minimal Supersymmetric extension of the Standard Model (MSSM), the tree-level Higgs quartic couplings are fixed to be the gauge couplings which leads to the tree-level Higgs boson mass that is below that of the $Z$ boson.
 Therefore, raising the Higgs boson mass to the observed value of 125~GeV at the LHC~\cite{Aad:2012tfa,Chatrchyan:2012ufa} requires large corrections due to a heavy stop sector~\cite{Okada:1990vk,Haber:1990aw,Ellis:1990nz,Barbieri:1990ja,
 Casas:1994us,Carena:1995bx,
 Carena:1995wu,Haber:1996fp,Heinemeyer:1998np,Carena:2000dp,Martin:2002wn}.
 However, heavy stops lead to large corrections to the up-type soft SUSY breaking Higgs squared mass parameter which in turn leads to a large correction to the electroweak symmetry breaking (EWSB) condition.
 A delicate cancellation between these corrections and the Higgsino mass parameter is needed to stabilize the electroweak scale, which is generally considered unnatural.
 Therefore in the MSSM there exists a tension between the observed Higgs mass and the requirement that the model is natural.

The Next-to-Minimal Supersymmetric Standard Model (NMSSM) is the simplest extension of the MSSM that can address this tension.
In the NMSSM, the Higgs sector is enlarged to include an extra gauge singlet that couples to the remaining MSSM Higgs doublets through a Yukawa coupling $\lambda$.
$\lambda$ contributes to the Higgs quartic at tree-level, and for large enough values, can raise the Higgs mass to the observed 125~GeV.
Therefore the stops need not be too heavy, thereby improving the naturalness of the model.
Moreover, in the general NMSSM (GNMSSM), an additional tadpole term for the gauge singlet can also facilitate EWSB~\cite{Ross:2011xv}.

For $\lambda \gsim 0.7$ at the weak scale, renormalization group (RG) evolution usually leads to this coupling
developing a Landau pole below the GUT scale. Refs.~\cite{Harnik:2003rs,
Chang:2004db,
Delgado:2005fq,
Craig:2011ev,
Hardy:2012ef,
Randall:2012dm}
have provided explicit UV-completions for such low scale models, which we collectively call Fat Higgs models.
Refs.~\cite{Barbieri:2006bg,
Cao:2008un,
Franceschini:2010qz,
Hall:2011aa,
Kanemura:2012uy,
Perelstein:2012qg,
Kyae:2012df,
Gherghetta:2012gb,
Barbieri:2013hxa,
Farina:2013fsa,
Gherghetta:2014xea,
Zheng:2014loa,
Cao:2014kya}
have studied the phenomenological implications of models with such large $\lambda$ couplings, which we collectively call $\lambda$-SUSY models.
For these models, they have found that the Higgs mass can easily be raised to the observed value while still keeping the spectrum natural.
These studies have focused on a region of low $\tan \beta$ ($\equiv v_u/v_d$, where $v_u$ and $v_d$ are the vacuum expectation values (VEVs) of the corresponding Higgs doublets) and large $\lambda$ because these regions were the most natural.

In this paper, we study the possibility of raising the Higgs mass to 125 GeV in Fat Higgs/$\lambda$-SUSY models at {\em large} $\tan \beta$.
As the $\lambda^2$-proportional tree-level contribution to the Higgs quartic is suppressed at large $\tan \beta$,
the one-loop induced radiative corrections are crucial in raising the Higgs boson mass to its observed value.
Similar to the stop-induced corrections that are proportional to $y_t^4 \log (m_{\tilde t}^2/Q^2)$ (where $y_t$ is the top Yukawa and $m_{\tilde t}$ is the stop mass scale), in Fat Higgs/$\lambda$-SUSY models the dominant one-loop corrections are proportional to $\lambda^4 \log (M_A^2/Q^2)$ (where $M_A$ is the scale of the non-standard Higgs bosons).  Therefore these corrections are only relevant when $\lambda \gsim 1$ and the non-standard Higgs bosons are much heavier than the electroweak scale.
The effect of radiative corrections in the NMSSM Higgs sector have been considered before~\cite{Gherghetta:2012gb,Ellwanger:2005fh}.
Ref.~\cite{Gherghetta:2012gb} focused on the most natural regions in the
Scale-Invariant NMSSM, where it was found that these radiative corrections
made a negligible contribution.
In contrast, we show that at large $\tan \beta$, the $\lambda$ induced radiative corrections can significantly modify the allowed regions of parameter space.
 Unlike Refs.~\cite{Barbieri:2006bg,Gherghetta:2012gb,Barbieri:2013hxa}, we also emphasize that electroweak precision constraints do not put a limit on $\tan \beta$.
We point out that raising the Higgsino mass parameter $\mue$ significantly weakens the electroweak precision constraints because the Higgsino component in the lightest neutralino is suppressed.
The price of raising $\mue$ is a slight increase in the tuning of the EWSB condition.
To illustrate these effects in regions of large $\tan \beta$ we also impose constraints from Higgs decay properties, direct  dark matter detection experiments, the observed dark matter relic density and the invisible width of the $Z$ boson.
In particular, we find that direct dark matter detection experiments place
strong limits on many regions of parameter space due to the large $\lambda$ coupling.
We also show that these allowed pockets of parameter space are within the reach of the XENON 1T experiment~\cite{Aprile:xenon1T}.

This paper is organized as follows.
In Sec.~\ref{sec:theory_setup}, we set up the theoretical aspects required for the phenomenology of our model.
To motivate the sizes of various terms in the Fat Higgs/$\lambda$-SUSY model,  we present a ``toy" high scale model where the fields have canonical mass dimensions in the electric theory.
In addition, in this section we also compute the corrections to the Higgs quartic using the one-loop effective potential formalism, and discuss the Higgsino contributions to electroweak precision constraints and naturalness in the large $\tan \beta$ regime of the Fat Higgs/$\lambda$-SUSY model.
In Sec.~\ref{sec:pheno}, we illustrate the impact of the formalism in Sec.~\ref{sec:theory_setup} by finding phenomenologically viable scenarios that can be probed at future experiments.
In Sec.~\ref{sec:conclusion} we conclude.

%SSSSSSSSSSSSSSSSSSSSSSSSSSSSSSSSSSSSSSSSSS
\section{Theoretical Setup}
\label{sec:theory_setup}

In this section we first motivate the form taken by our superpotential by a simple discussion of the sizes of various terms that can arise in Fat Higgs/$\lambda$-SUSY models.
In this discussion we assume that any exotic fields are much heavier than the electroweak scale.
For the superpotential thus obtained, we present the Higgs potential at the tree level and analytically compute the one-loop corrections to the mass of the SM-like Higgs boson due to heavy non-standard Higgs fields,
with special attention to the limit of large $\lambda$ and $\tan \beta$.
In addition we discuss the naturalness of the large $\tan\beta$ regions of the Fat Higgs/$\lambda$-SUSY models.
We then discuss the reduced couplings of the SM-like Higgs to SM particles, which are constrained by LHC measurements of signal strengths.
We end the section with a brief discussion of the neutralino sector with particular attention to electroweak precision observables.

%SSSSSSSSSSSSSSSSSSSSSSSSSSSSSSSSSSSSSSSSSS
\subsection{Realizing low scale NMSSM with large $\lambda$}
\label{sec:lowscaleNMSSM}

The GNMSSM with a large $\lambda$ at the weak scale implies that some of the Higgs fields are composite states.
For example, in the minimal Fat Higgs scenario of Ref.~\cite{Harnik:2003rs}, all of the Higgs sector fields are composite, while in Refs.~\cite{Chang:2004db,Randall:2012dm} the MSSM Higgs fields are fundamental.
For simplicity we will assume that at scales $\lsim 10$~TeV, the only fields present in the Higgs sector are the $SU(2)_L$ doublets $\hat H_u,\hat H_d$ and the gauge singlet $\hat S$.\footnote{More exotic realizations typically require that additional Standard Model superfields are composite, introducing more exotic fields in the low energy theory~\cite{Delgado:2005fq,Craig:2011ev,Csaki:2011xn,Csaki:2012fh}.}

The most general superpotential with this particle content (assuming R-parity) has the form~\cite{Ellwanger:2009dp,Ross:2011xv}
%%%%%%%%%%%%%%%%%%%%%%%%%%%%%%%
\bea
\mathcal{W}_{\rm GNMSSM} = \mathcal{W}_{\rm Yukawa} +  \lambda \hat S \hat H_u \hat H_d  + \frac{1}{3} \kappa \hat S^3 + \muH \hat H_u \hat H_d + \frac{1}{2} \mu' \hat S^2 + \xi_F \hat S,
 \label{eq:NMSSMsuperpotential}
\eea
%%%%%%%%%%%%%%%%%%%%%%%%%%%%%%%
where $\lambda, \kappa$ are dimensionless coupling strengths; $\muH, \mu'$ are supersymmetric mass terms; $\xi_F$ is a supersymmetric tadpole term of mass dimension 2, and $\mathcal{W}_{\rm Yukawa}$ contains the standard MSSM Yukawa superpotential terms. The corresponding soft SUSY-breaking terms are
%%%%%%%%%%%%%%%%%%%%%%%%%%%%%%%
\bea
-\mathcal{L}_{\rm soft} &=& -\mathcal{L}_{\rm soft}^{\tilde f} + m_{H_u}^2 |H_u|^2 + m_{H_d}^2 |H_d|^2 + m_{S}^2 |S|^2 + \nonumber \\
& & \left(\lambda A_\lambda H_u H_d S + \frac{1}{3} \kappa A_\kappa S^3 + m_3^2 H_u H_d + \frac{1}{2} {m'_S}^{2} S^2 + \xi_S S + h.c.\right),
 \label{eq:NMSSMsoft}
\eea
%%%%%%%%%%%%%%%%%%%%%%%%%%%%%%%
where $\mathcal{L}_{\rm soft}^{\tilde f}$ corresponds to the standard MSSM soft
SUSY-breaking terms.
$m_{H_u}^2, m_{H_d}^2, m_S^2$ are the soft SUSY breaking Higgs squared mass terms and $A_\lambda, A_\kappa$ are the soft SUSY breaking trilinear terms.
$m_3^2, {m'_S}^2$  are the CP-violating soft SUSY breaking squared mass terms and $\xi_S$ is the dimension-3 soft SUSY breaking term corresponding to $\xi_F$.

A generic feature of most Fat Higgs/$\lambda$-SUSY models is that the Yukawa coupling $\lambda \gsim 0.7$ at the TeV scale.\footnote{For Fat Higgs models that provide an existence proof of gauge coupling unification, see \cite{Harnik:2003rs,Chang:2004db}.}
Due to its renormalization group (RG) evolution, $\lambda$ becomes stronger at higher scales and develops a Landau pole at the compositeness scale $\Lambda_H$, where $\Lambda_H$ is assumed to be lower than the grand unification scale $M_{\rm GUT}$.
In the deep IR, much below $\Lambda_H$, the magnetic theory of mesons (i.e. the Higgs superfields) is described by the interactions in Eqs.~(\ref{eq:NMSSMsuperpotential}) and (\ref{eq:NMSSMsoft}).
In the UV above $\Lambda_H$,  some or all of the Higgs superfields are revealed to be composite states made up fundamental quarks whose interactions are described by some electric theory.

If the quarks in the electric theory have the canonical mass dimension and all Higgs superfields are composite, then the $\kappa, \muH, \mu'$ terms in Eq.~(\ref{eq:NMSSMsuperpotential}) and their corresponding soft SUSY-breaking terms in Eq.~(\ref{eq:NMSSMsoft}) are generated by marginal terms in the fundamental theory.
This assumption is equivalent to saying that the confining dynamics only generates the $\lambda$ term while all other couplings arise from non-renormalizable interactions in the electric theory.
For example, in the simplest Fat Higgs model~\cite{Harnik:2003rs}, the Higgs superfields in Eq.~(\ref{eq:NMSSMsuperpotential}) are composite states of the quarks $T_i$ in the electric theory.
These quarks are charged under a confining $SU(2)_H$ gauge group, thereby leading to the identification
%%%%%%%%%%%%%%%%%%%%%%%%%%%%%%%%%%%%%%%%%
\bea
\hat S \sim \hat T_5 \hat T_6; \;\; \bmat{c} \hat H_u^+ \\ \hat H_u^0 \emat \sim \bmat{c} \hat T_1 \hat T_3 \\ \hat T_2 \hat T_3 \emat; \;\;
\bmat{c} \hat H_d^0 \\ \hat H_d^- \emat \sim \bmat{c} \hat T_1 \hat T_4 \\ \hat T_2 \hat T_4 \emat. \label{eq:fathiggscomposites}
\eea
%%%%%%%%%%%%%%%%%%%%%%%%%%%%%%%%%%%%%%%%%%
The $\lambda$ term in Eq.~(\ref{eq:NMSSMsuperpotential}) is dynamically generated by the Pfaffian of the mesons in the magnetic theory.
 Naive dimensional analysis (NDA) \cite{Manohar:1983md,Luty:1997fk,Cohen:1997rt} and canonical normalization of the fields $\left(\langle T_i T_j \rangle \to (\Lambda_H/4\pi) \phi_{ij}\right)$ lead to the relations
%%%%%%%%%%%%%%%%%%%%%%%%%%%%%%%%%%%%%%%%
\bea
\lambda(\Lambda_H) \sim 4\pi; &\;\;\;& \kappa(\Lambda_H) \sim \left(\frac{\Lambda_H}{4\pi \Lambda_0}\right)^3 \nonumber \\
\muH \sim \frac{\Lambda_H^2}{(4\pi)^2 \Lambda_0} \sim \mu' ; &\;\;\;&
\xi_F \sim \frac{m \Lambda_H}{4\pi},
\label{eq:NDAestimates}
\eea
%%%%%%%%%%%%%%%%%%%%%%%%%%%%%%%%%%%%%%
where $m$ and $\Lambda_0$ are parameters in the electric superpotential given by
%%%%%%%%%%%%%%%%%%%%%%%%%%%%%%%%%%%%%%
\bea
\mathcal{W}_{\rm electric} &\simeq& m \hat T_5 \hat T_6 + \frac{y}{\Lambda_0} \left(\hat T_5 \hat T_6 \right)^2 + \frac{y'}{\Lambda_0} \left[\left(\hat T_1 \hat T_3 \right) \left(\hat T_2 \hat T_4 \right) - \left(\hat T_1 \hat T_4 \right) \left(\hat T_2 \hat T_3 \right) \right]+ \nonumber \\ &&\frac{y''}{\Lambda_0^3} (T_5 T_6)^3.
\eea
%%%%%%%%%%%%%%%%%%%%%%%%%%%%%%%%%%%%%
The couplings $y, y',y''$, in the above equation, need not be $\mathcal{O}(1)$ numbers because $\Lambda_0$ is just a generic scale used to parameterize the mass dimension of each of these operators.

Eq.~\ref{eq:NDAestimates} gives us a definition of $\Lambda_H$: it is the scale at
which the size of $\lambda$ is $4\pi$.
Using this definition, we can then estimate the size of the other
parameters at the weak scale from their RG evolution.
In determining $\Lambda_H$,
we also account for the effects of the SM Yukawa couplings
using the renormalization group equations (RGEs) in Ref.~\cite{Ellwanger:2009dp}.
Having estimated the NMSSM parameters at the scale $\Lambda_H$ using Eq.~\ref{eq:NDAestimates},
we run them down to the TeV scale by solving the RGEs and find that they \textit{decrease} with decreasing scale.
This running behavior has two important implications for our model:

%%%%%%%%%%%%%%%%%%%%%%%%%%%%5
% \bea
% \frac{d\lambda}{d \log \mu} &=& \frac{\lambda}{16\pi^2} (4\lambda^2 + 2\kappa^2 + 3y^2_t + 3y^2_b + 3y^2_\tau), \nonumber \\
% \frac{d\kappa}{d \log \mu} &=& \frac{\kappa}{16\pi^2} (6\lambda^2 + 6\kappa^2),  \nonumber \\
% \frac{dy_t}{d \log \mu} &=& \frac{y_t}{16\pi^2} (\lambda^2 + 6y^2_t + y^2_b), \nonumber \\
% \frac{dy_b}{d \log \mu} &=& \frac{y_b}{16\pi^2} (\lambda^2 + y^2_t + 6y^2_b + y^2_\tau),  \nonumber \\
% \frac{dy_\tau}{d \log \mu} &=& \frac{y_\tau}{16\pi^2} (\lambda^2 + 3y^2_b + 4y^2_\tau),  \nonumber \\
% \frac{d\muH}{d \log \mu} &=& \frac{\muH}{16\pi^2} (2\lambda^2 + 3y^2_t + 3y^2_b+y^2_\tau),  \nonumber \\
% \frac{d\mu'}{d \log \mu} &=& \frac{\mu'}{16\pi^2} (4\lambda^2 + 4\kappa^2).  \nonumber \\
% \label{eq:RGEs}
% \eea
%%%%%%%%%%%%%%%%%%%%%%%%%%%%

%We have neglected the contribution of the gauge couplings in the above.

%%%%%%%%%%%%%%%%%%%%%%%%%%%%%%
\begin{figure}
\begin{center}
\includegraphics[width=0.48\textwidth]{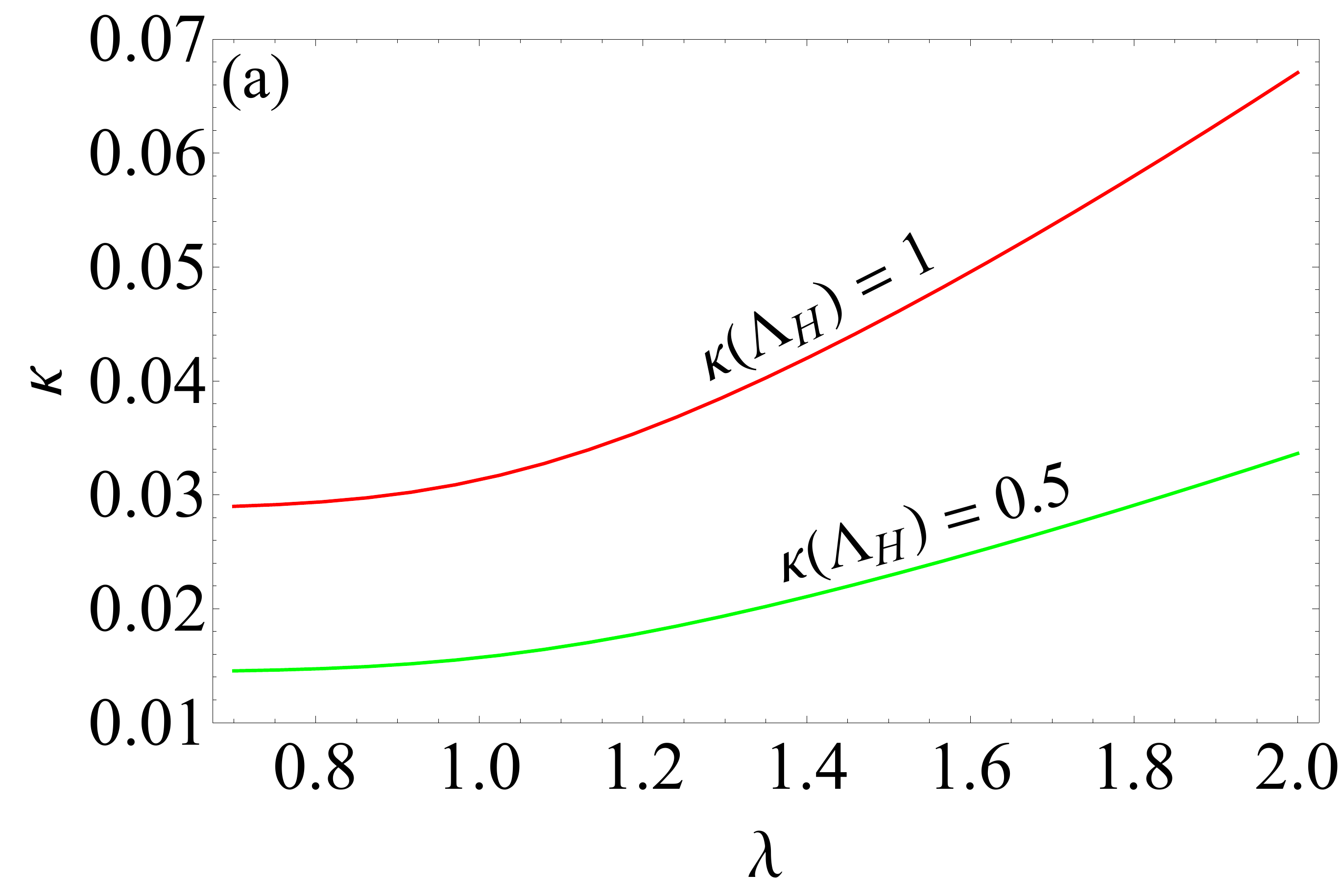}
\includegraphics[width=0.48\textwidth]{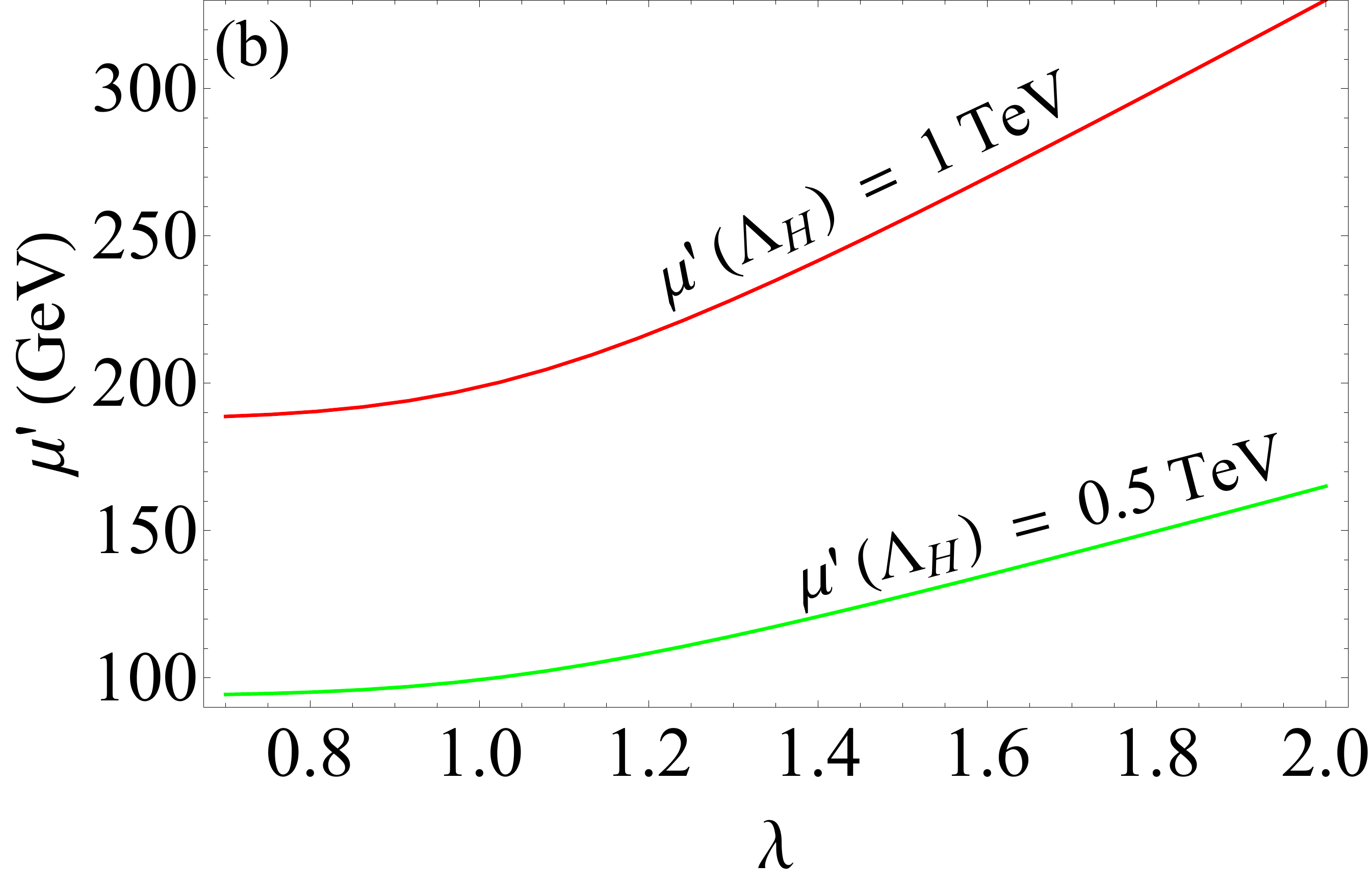}
\caption{\footnotesize (a): $\kappa$ as a function of $\lambda$ at the scale $Q=M_Z$, obtained by
fixing $\kappa$ at the scale $\Lambda_H$ and then evolving it down with RGEs.
The red (green) curve corresponds to $\kappa(\Lambda_H) = 1 (0.5)$.
(b): $\mu'$ as a function of $\lambda$ at the scale $Q=M_Z$, obtained in a manner
analogous to (a).
The red (green) curve corresponds to $\mu'(\Lambda_H) = 1 (0.5)~\tev$.
In both plots we set $\tan \beta = 50$.
See text for details of their behavior.}
\label{fig:kapparun}
\end{center}
\end{figure}
%%%%%%%%%%%%%%%%%%%%%%%%%%%%%%

1. Eq.~(\ref{eq:NDAestimates}) implies $\kappa(\Lambda_H) \ll \mathcal{O}(1)$.
  Run down to a renormalization scale $Q = \mathcal{O}(\tev)$, we expect $\kappa$ to be quickly suppressed due to the contribution of $\lambda$ to its running.
 This suppression is illustrated in Fig.~\ref{fig:kapparun}(a), where we plot $\kappa$ at $Q = M_Z$ as a function of $\lambda$ at $Q = M_Z$, setting $\tan \beta = 50$.
 These curves were obtained
 by first running $\lambda(Q = M_Z)$ up to determine $\Lambda_H$,
 then setting $\kappa (Q = \Lambda_H)$ to different values $\leq 1$,
 and finally running $\kappa$ down to $Q = M_Z$.
 We checked that the running of $\lambda$ is insensitive to $\kappa$ for
 these sizes of $\kappa$.
 The red curve corresponds to $\kappa (\Lambda_H) = 1$ and the green curve to $\kappa (\Lambda_H) = 0.5$.
 As expected from the RG running, smaller values of $\kappa (\Lambda_H)$ result in smaller values of $\kappa (Q = M_Z)$.

 A larger $\lambda$ implies a Landau pole at a lower scale.
 Therefore, $\Lambda_H$ is closer to the electroweak scale for larger
 values of $\lambda$, which in turn weakens the
suppression of $\kappa$ as it runs down from $\Lambda_H$ to $M_Z$.
This is why $\kappa$ is an increasing function of $\lambda$ in Fig.~\ref{fig:kapparun}(a).
From the plot, we infer that for $\kappa (\Lambda_H) \leq 1$,
the size of $\kappa$ at the weak scale is suppressed by at least an order of magnitude.
The implication of this suppression is that we can consistently neglect the effects of $\kappa$ in our TeV-scale phenomenology.
Therefore, for the rest of this paper we will take $\kappa = 0$.

2. As compared to $\kappa$, $\mu'$ is only suppressed by an $\mathcal{O}(1)$ number when it is run down from $Q=\Lambda_H$ and $Q=M_Z$. This difference between
values of $\mu'$ and $\kappa$ can be understood from their $\beta$-function
dependences. Using their one-loop $\beta$-functions in
Ref.~\cite{Ellwanger:2009dp} we find
\bea
\frac{\kappa(Q)}{\kappa(\Lambda_H)} = \left(\frac{\mu'(Q)}{\mu'(\Lambda_H)}\right)^3.
\eea
%This is important because we require $60~\gev~\lsim\mu'~\lsim~1~\tev$ for locating phenomenologically viable regions of our model, and therefore an RG-suppression that is either too slow or too rapid may put us in tension with the estimates  in Eq.~\ref{eq:NDAestimates}.
We check this by determining $\mu' (Q=M_Z)$ as a function of $\lambda (Q=M_Z)$ in a manner analogous to the determination of $\kappa (Q=M_Z)$ above. Our results are shown in Fig.~\ref{fig:kapparun}(b), where the red (green) curve corresponds to $\mu' (\Lambda_H) = 1(0.5)~\tev$, with $\tan \beta = 50$.  We see that $\mu' (\Lambda_H)$ is suppressed at the electroweak scale by at most a factor of 5. Hence $\mu' (M_Z)$ can be of the size of the electoweak scale. 
Such a size results from reasonable values of the scale $\Lambda_0$.
For instance, to obtain $\mu' (Q=M_Z) \lsim 1~\tev$ at $\tan \beta = 50$, we computed from Eq.~\ref{eq:NDAestimates}
that $\Lambda_0 \leq 10^{16}~\gev$ for $\lambda (Q=M_Z) \geq 0.7$. 
Similarly, the Higgsino mass parameter $\mu$ and the tadpole term $\xi_F^{1/2}$ can also be the size of the electroweak scale.

We can now write down our low energy superpotential below the scale $\Lambda_H$:
%%%%%%%%%%%%%%%%%%%%%%%%%%%%%%%%%%%%%%%%%%%%%%%%%%%%
\bea
\mathcal{W}_{\rm NMSSM}^{\rm eff} = \mathcal{W}_{\rm Yukawa} +  \lambda \hat S \hat H_u \hat H_d  + \frac{1}{2}\mu' \hat S^2 + \xi_F \hat S
 \label{eq:NMSSMeff}
\eea
%%%%%%%%%%%%%%%%%%%%%%%%%%%%%%%%%%%%%%%%%%%%%%%%%%%%
The associated soft-SUSY breaking potential is
%%%%%%%%%%%%%%%%%%%%%%%%%%%%%%%%%%%%%%%%%%%%%%%%%%%%
\bea
-\mathcal{L}_{\rm soft}^{\rm eff} &=& -\mathcal{L}_{\rm soft}^{\tilde f} + m_{H_u}^2 |H_u|^2 + m_{H_d}^2 |H_d|^2 + m_{S}^2 |S|^2 \nonumber \\
& & + \left(\lambda A_\lambda S H_u H_d  + m_3^2 H_u H_d + \frac{1}{2} m'^2_S + \xi_S S + {\rm h.c.}\right). \label{eq:NMSSMsofteff}
\eea
%%%%%%%%%%%%%%%%%%%%%%%%%%%%%%%%%%%%%%%%%%%%%%%%%%%
  We have redefined the singlet chiral superfield, $\hat S \to \hat S - \muH$, to remove the $\muH$ term in the superpotential. In general, the associated soft term $m_3^2$ cannot be absorbed into $A_\lambda$ simultaneously.
  Eqs.~(\ref{eq:NMSSMeff}) and Eq.~(\ref{eq:NMSSMsofteff}) constitute all the parameters treated in the rest of this article.

%%%%%%%%%%%%%%%%%%%%%%%%%%%%%%%%%%%%%%%%%%%%%%%%%%
\subsection{Higgs Sector}
\label{subsec:higgs_sector}
\subsubsection{Tree level}
At the tree level, the Higgs potential is given by
%%%%%%%%%%%%%%%%%%%%%%%%%%%%%%%%%%%%%%%%%%%
\bea
V_{\rm Higgs}^{\rm tree} = V_F  + V_D + V_S %+ \frac{1}{16 \pi^2} \Delta V_{1-\rm loop},
\label{eq:Higgspottree}
\eea
%%%%%%%%%%%%%%%%%%%%%%%%%%%%%%%%%%%%%%%%%%%
where
%%%%%%%%%%%%%%%%%%%%%%%%%%%%%%%%%%%%%%%%%%%
\bea
V_F &=&   \left| \lambda \left(H_u^+ H_d^- - H_u^0 H_d^0\right) + \mu' S + \xi_F \right|^2 +  \left|\lambda S\right|^2\ (\left|H_u\right|^2 + \left|H_d\right|^2 ), \nonumber \\
V_D &=& \frac{g^2}{8} \left(\left|H_u^0\right|^2 + \left|H_u^+\right|^2 - \left|H_d^0\right|^2 - \left|H_d^-\right|^2 \right) + \frac{g^2}{2}\cos^2\theta_W \left|H_u^+ H_d^{0*} + H_u^0 H_d^{-*}\right|^2, \label{eq:higgspot_allterms} \\
V_S &=&  m_{H_u}^2 |H_u|^2 + m_{H_d}^2 |H_d|^2 + m_S^2 |S|^2  + \nonumber \\
& & (\lambda A_\lambda(H_u^+ H_d^- - H_u^0 H_d^0) S + m_3^2 (H_u^+ H_d^- - H_u^0 H_d^0) + \frac{1}{2} m^{'2}_S S^2 + \xi_S S + {\rm h.c.}), \nonumber
%V_{1-\rm loop} &=& \frac{1}{64\pi^2} \left(\left(\frac{1}{2} \sum_{i=1}^{3}  \left(M_H^2\right)_i^2 \log \left[\frac{\left(M_H^2\right)_i}{Q^2}\right]\right) + \frac{1}{2} \sum_{i=1}^{3} \left(M_A^2\right)_i^2 \log \left[\frac{\left(M_A^2\right)_i}{Q^2}\right]\right)
\eea
%%%%%%%%%%%%%%%%%%%%%%%%%%%%%%%%%%%%%%%%%%%%
$H_u = (H_u^+, H_u^0)$, $H_d = (H_d^0, H_d^-),~g^2 \equiv g^2_1 + g^2_2$ and $\theta_W$ is the weak mixing angle.
After electroweak symmetry breaking we can expand the Higgs fields in terms of the CP-even fields $(h_u^0, h_d^0, h_s^0)$, the CP-odd fields $(A_D^0,A_S^0)$, the charged Higgs bosons $H^\pm$ and the Goldstone bosons $(G^\pm, G^0)$:
%%%%%%%%%%%%%%%%%%
\bea
H_u &=& \left( \begin{array}{c}
    G^+s_\beta + H^+c_\beta      \\
    v s_\beta + \frac{1}{\sqrt{2}} [ (h_u^0 + i (G^0 s_\beta - A_D^0 c_\beta)  ]
       \end{array} \right),
\nonumber \\
H_d &=& \left( \begin{array}{c}
    v c_\beta + \frac{1}{\sqrt{2}} [(h_d^0 + i (- G^0 c_\beta + A_D^0 s_\beta)  ]  \\
       - G^- c_\beta + H^- s_\beta
       \end{array} \right),
\nonumber \\
S &=&  \frac{1}{\sqrt{2}}(s+ h^0_s + i A^0_S),
\label{eq:higgsafterEWSB}
\eea
%%%%%%%%%%%%%%%%%%
where $v \simeq 174~\gev$ is the VEV of EWSB, $s_\beta \equiv \sin \beta,~c_\beta \equiv \cos \beta$ and $s \equiv \langle S \rangle$.
Expanding the potential about the minimum at $v_i \equiv (v_u,v_d,s)$, we can find the tree-level tadpole terms
%%%%%%%%%%%%%%%%%%%%%%%%%%%%%%%%%%%%%%%%%%%%%%%%%%%
\bea
% T_u^{\rm tree} \equiv \left. \frac{\partial V_{\rm Higgs}^{\rm tree}}{\partial H_u^0} \right|_{\{v_i\}} &=& v_u \left(m_{H_u}^2 + \mu_{\rm eff}^2 + \lambda^2 v_d^2  + \frac{g_1^2 + g_2^2}{4}(v_u^2 - v_d^2)\right) \nonumber\\
%&-& v_d \left(\mu_{\rm eff} (A_\lambda+\mu') + m_3^2 + \lambda \xi_F \right) \nonumber \\
%%%%%%%%%%%
% T_d^{\rm tree} \equiv \left. \frac{\partial V_{\rm Higgs}^{\rm tree}}{\partial H_d^0} \right|_{\{v_i\}} &=& v_d \left(m_{H_u}^2 + \mu_{\rm eff}^2 + \lambda^2 v_u^2  + \frac{g_1^2 + g_2^2}{4}(v_d^2 - v_u^2)\right)  \nonumber\\
%&-& v_u \left(\mu_{\rm eff} (A_\lambda+\mu') +  m_3^2 + \lambda \xi_F \right) \nonumber  \\
%%%%%%%%%%
%T_s^{\rm tree}\equiv \left. \frac{\partial V_{\rm Higgs}^{\rm tree}}{\partial S} \right|_{\{v_i\}} &=& s \left(m_S^2 + {m'}^2_S + {\mu'}^2 \right) + \lambda^2(v_u^2 + v_d^2) + \xi_s + \xi_F \mu' \nonumber \\
%&-& \lambda v_u v_d (A_\lambda + \mu')
T_j^{\rm tree}\equiv \left. \frac{\partial V_{\rm Higgs}^{\rm tree}}{\partial \phi_j} \right|_{\{v_i\}}
\label{eq:tadpoles_tree}
\eea
%%%%%%%%%%%%%%%%%%%%%%%%%%%%%%%%%%%%%%%%%%%%%%%%%%%%
where $\phi_j = (H_u^0,H_d^0,S)$.
We can then solve for the soft squared masses $m_{H_u}^2,m_{H_d}^2,m_S^2$ by setting each $T_j^{\rm tree}  = 0$.
Substituting these masses into the second order derivatives of the Higgs potential and neglecting CP-violating effects,
we obtain the following tree-level CP-even Higgs mass matrix in the basis $(h^0_u,h^0_d,h^0_s)$.
%%%%%%%%%%%%%%%%%%%%%%%%%%%%%%
\bea
\left(M^2_H\right)_{11} = M^2_Z s^2_\beta + r~t^{-1}_\beta; &\;\;\;\;\;&
\left(M^2_H\right)_{12} = (2\lambda^2 v^2 - M^2_Z) s_\beta c_\beta - r; \nonumber \\
\left(M^2_H\right)_{22} = M^2_Z c^2_\beta + r~t_\beta; &\;\;\;\;\;\;&
\left(M^2_H\right)_{13} =  \lambda v(2\mue s_\beta - (A_\lambda + \mu') c_\beta);   \label{eq:massmatcpeven} \\
\left(M^2_H\right)_{23} = \lambda v(2\mue c_\beta - (A_\lambda + \mu') s_\beta); &\;\;\;\;\;\;&
\left(M^2_H\right)_{33} = \left( \lambda v^2 (A_\lambda + \mu') - (\xi_S + \xi_F \mu') \right)/s,  \nonumber
\eea
%%%%%%%%%%%%%%%%%%%%%%%%%%%%%%
where $\mue \equiv \lambda s,~t_\beta \equiv \tan\beta$ and $r \equiv \mue(A_\lambda+\mu') + m^2_3 + \lambda \xi_F$.
The CP-odd Higgs mass matrix in the basis $(A_D^0,A_S^0)$ is given by
%%%%%%%%%%%%%%%%%%%%%%%%%%%%%%%%%%%%%%%
\bea
&\left(M^2_A\right)_{11} = 2 r/ s_{2\beta}; \;\;\;\;\;\; \left(M^2_A\right)_{12} =  \lambda v (\Al - \mu'); & \nulein
 &\left(M^2_A\right)_{22}  = \frac{1}{s} \left(\lambda v^2 (\Al + \mu') s_\beta c_\beta - (\xi_F \mu' + \xi_S)\right)  - 2{m'}_S^2,
\label{eq:massmatcpodd}
\eea
%%%%%%%%%%%%%%%%%%%%%%%%%%%%%%%%%%%%%%
and the charged Higgs mass is
%%%%%%%%%%%%%%%%%%%%%%%%%%%%%%%%%%%%%%
\bea
M^2_{\pm} = 2r/s_{2\beta} - (\lambda^2 - g^2_2/2) v^2. \label{eq:massmatcharged}
\eea
%%%%%%%%%%%%%%%%%%%%%%%%%%%%%%%%%%%%%%

We point out two features of the tree level masses that will be important in our discussion of the one-loop corrected Higgs mass.
The first feature is the correlation among the scalar masses in the limit where $\Al$ and $\mu'$ are small compared to the heavy
Higgs masses.
This is best seen by setting $\left(M^2_A\right)_{12} = 0$ in Eq.~(\ref{eq:massmatcpodd}) (which can be
obtained by choosing $\Al = \mu'$).
Then the CP-odd eigenmasses are identified as $\MAonesq = \left(M^2_A\right)_{11}$ and $\MAtwosq = \left(M^2_A\right)_{22}$.
In this limit, by inspecting the matrix elements in Eqs.~(\ref{eq:massmatcpeven})--(\ref{eq:massmatcharged}),
we find that the CP-even, CP-odd and charged Higgs eigenstates arising from the $SU(2)$ doublet sector are nearly degenerate in mass, a feature well-known in the MSSM.
Their mass splittings $\sim v^2$.
These three fields then have a mass $\sim M_{A_D}$ in the limit $\MAonesq \gg v^2, \Al^2,\mu'^2$, where $M_{A_D}$ denotes the corresponding CP-odd eigenmass.
Likewise, the CP-even and CP-odd Higgs eigenstates arising from the $SU(2)$ singlet are nearly degenerate, with mass splitting $\sim s^2$.
Therefore, these two fields have a mass $\sim M_{A_S}$ in the limit $\MAtwosq \gg s^2,\Al^2,\mu'^2$.

 The second feature is the decoupling of heavy states.
 Raising $\MAonesq$ and $\MAtwosq$ decreases their impact on the mass of the lightest CP-even state, effectively
 making it more SM-like.
 A simple way to see this decoupling behavior is to rotate the CP-even mass matrix into the basis
 %%%%%%%%%%%%%%%
\bea
h^0 = h_u^0 s_\beta + h_d^0 c_\beta,~ H^0 = h_u^0 c_\beta - h_d^0 s_\beta,~h^0_s = h^0_s \label{eq:bogdanbasis}
\eea
%%%%%%%%%%%%%%%
which leads to the CP-even mass matrix
%%%%%%%%%%%%%%%%%%%%%%%%%%%%%%
\bea
\left(M^2_H\right)_{hh} =   M^2_Z c^2_{2\beta} + \lambda^2 v^2 s^2_{2\beta} ; &\;\;\;\;\;&
\left(M^2_H\right)_{hH} = (\lambda^2v^2-M_Z^2) s_{4\beta}/2 ; \nonumber \\
\left(M^2_H\right)_{HH} = \MAonesq - (\lambda^2v^2-M_Z^2) s^2_{2\beta} ; &\;\;\;\;\;\;&
\left(M^2_H\right)_{hS} =  2\lambda v (\mue - \Al s_{2\beta}) ; \label{eq:decoupling_tree_Bogdan} \\
\left(M^2_H\right)_{HS} =  -2 \lambda v \Al c_{2\beta} ; &\;\;\;\;\;\;&
\left(M^2_H\right)_{SS} = \MAtwosq + 2 {m'_S}^2 + {\lambda^2 v^2 \over \mue} \Al (2-s_{2\beta} c_\beta),  \nonumber
\eea
%%%%%%%%%%%%%%%%%%%%%%%%%%%%%
Notice that $\xi_F, \xi_S$ and $m^2_3$ are absorbed into our definition of
$\MAonesq$ and $\MAtwosq$.
For large $\tan \beta$, $\MAonesq$ and $\MAtwosq$, $h^0$ is identified with the SM Higgs,
 and $H^0$ and $h^0_s$ with non-standard Higgs bosons.
 This decoupling feature should be preserved after the inclusion of radiative corrections to the lightest CP-even Higgs mass,
 which is a non-trivial check of this computation.

%In the limit of large masses the splittings between these states becomes negligible.
%In particular, if  $ \left(M^2_A\right)_{12}$ is negligible then we can identify the mass CP-odd masses as $\MAonesq \equiv  \left(M^2_A\right)_{11}$ and $\MAtwosq \equiv  \left(M^2_A\right)_{22}$.
%For large $\MAonesq$, the mass splittings between the CP-odd, CP-even and charged Higgs bosons arising from the doublet sector become negligible. %Similarly for large $\MAtwosq$ the mass splittings between the CP-odd and CP-even Higgs bosons arising from the singlet sector also become negligible.

\subsubsection{Radiative corrections}
\label{subsec:radiatifcorrex}

The mass of the lightest Higgs boson can be significantly modified by one-loop corrections. The largest contributions to the Higgs potential at one-loop level are from the Higgs bosons, third generation squarks, charginos and neutralinos. Thus we have
%%%%%%%%%%%%%%%%%%%%%%%%%%%%%%%%%%%%%%%%%%%
\bea
\Delta V &=& \frac{1}{32\pi^2} \left(3 \Delta V^{\tilde t} -6 \Delta V^{t} - \Delta V^{\chi^\pm} -2 \Delta V^{\chi^0} + \frac{1}{2} \Delta V^{H} + \frac{1}{2} \Delta V^{A} +  \Delta V^{H^\pm}\right),
\label{eq:DeltaV_full}
\eea
%%%%%%%%%%%%%%%%%%%%%%%%%%%%%%%%%%%%%%%%%%%
where for the $a^{\rm th}$ sector in the $\overline{\text{MS}}$ scheme,
%%%%%%%%%%%%%%%%%%%%%%%%%%%%%%%%%%%%%%%%%%%
\bea
\Delta V^a =  \sum_i \left(M_{ia}^2(\phi_k)\right)^2 \left(\log\frac{M_{ia}^2(\phi_k)}{Q^2} - \frac{3}{2}\right) \approx \sum_{i =\rm heavy} \left(M_{ia}^2(\phi_k)\right)^2 \left(\log\frac{M_{ia}^2}{Q^2} - \frac{3}{2}\right).
\label{eq:effpoti}
\eea
%%%%%%%%%%%%%%%%%%%%%%%%%%%%%%%%%%%%%%%%%%%
$M_{ia}^2(\phi_k)$ is the field-dependent mass eigenvalue for $i^{\rm th}$ contribution, $M_{ia}^2$ is the corresponding field-independent tree-level eigenvalue
%, $\phi_k = (\phi_u,\phi_d,\phi_s)$ are the neutral fields corresponding to $(H_u^0, H_d^0,S)$ respectively
and the renormalization scale $Q \sim m_h = 125$~GeV. The approximation in Eq.~(\ref{eq:effpoti}) holds because we are interested in large corrections to the lightest Higgs mass due to states much heavier than the electroweak scale.  Also, the field dependences inside $\log$ terms are neglected since they only induce higher order field-dependent terms.

The dominant contributions to $\Delta V$ in our scenario are due to heavy Higgs scalars coupling to the light Higgs boson with strengths proportional to powers of $\lambda$.
The effects of the top quarks and the scalar tops on the Higgs potential have been studied in great detail in Refs.~\cite{Okada:1990vk,Haber:1990aw,Ellis:1990nz,Barbieri:1990ja,Casas:1994us,Carena:1995bx,Carena:1995wu,Haber:1996fp,Heinemeyer:1998np,Carena:2000dp,Martin:2002wn}. To highlight the effect of large $\lambda$ corrections, we suppress the contribution of scalar tops to $\Delta V$ by choosing their masses close to the electroweak scale while still being compatible with ATLAS and CMS bounds \cite{atlas_stops_semi,atlas_stops_had,cms_stops}.
The contributions of charginos and neutralinos to $\Delta V$ are typically small.
The Higgs couples to the bino and the wino triplet with electroweak strength, whereas the $\lambda$-dependent
coupling to the Higgsinos and singlino is typically suppressed due to neutralino mixing.
In addition, the masses of the Higgsinos and singlino $\lsim$ 1 TeV in our phenomenology while $M_A \in [4, 8]~\tev$.
We therefore neglect corrections from the chargino-neutralino sector in the remainder of this article.
% In our phenomenological discussions, the masses of charginos and neutralinos are also typically close to the electroweak scale;
% even in cases where they are taken heavy, they couple to the Higgs boson with electroweak strength. Thus, for the remainder of %the paper we will neglect their contributions to the Higgs mass.
% In our phenomenological discussions, the masses of charginos and neutralinos are also typically close to the electroweak scale;
% even in cases where they are taken heavy, they couple to the Higgs boson with electroweak strength. Thus, for the remainder of %the paper we will neglect their contributions to the Higgs mass.

In order to compute the one-loop corrections to the %lightest CP-even Higgs mass
Higgs potential in Eq.~(\ref{eq:effpoti}) due to the heavy CP-even, CP-odd and charged Higgs bosons, we must determine the field-dependent eigenvalues of each of the respective matrices.
When expressed in terms of the matrix elements these field-dependent eigenvalues can in general be quite complicated.
The calculation can nevertheless be simplified if we expand the eigenvalues as a Taylor series in the tree-level masses:
%(\textbf{NR: We write all (charged, CP-odd, CP-even) matrices as a series in $m^2_{A_i}$, as explained in the write-up in the appendix.})
%%%%%%%%%%%%%%%%%%%%%%%%%%%%%
\begin{equation}
M^2_i(\phi_k) =  M^2_{i,{\rm tree}} + \hat{b}_{i}(\phi_k) + \frac{ \hat{c}_i(\phi_k)}{M^2_{i,{\rm tree}}} + \mathcal{O}\left(\frac{1}{M^4_{i,{\rm tree}}}\right), \label{eq:fielddependentmass}
\end{equation}
%%%%%%%%%%%%%%%%%%%%%%%
%(\textbf{NR: The above equation says that an eigenvalue can be expanded as $M^2_i(\phi_k) = a_{i1} m^2_{A_1} + a_{i2} m^2_{A_2} + b_{i}(\phi_k) + c_{i1}(\phi_k) \frac{1}{m^2_{A_1}} + c_{i2}(\phi_k) \frac{1}{m^2_{A_2}}$; however, a given eigenvalue can be expanded only in terms of one pseudoscalar mass, not both. If we expand in both, we end up with the term $2a_{ij}c_{ik}\frac{m_{A_j}^2}{m_{A_k}^2}$ in Eq. (22) below.})
where the coefficients $\hat{b}_{i}(\phi_k)$ and $\hat{c}_{i}(\phi_k)$ are at most quadratic and quartic in the fields  respectively.
Furthermore, when evaluated at the tree-level VEVs, the coefficients satisfy the condition $\hat{b}(\{v_k\})=0=\hat{c}(\{v_k\})$.

In practice, we expand the eigenvalues as a Taylor series in the pseudoscalar masses $\MAonesq$ and $\MAtwosq$.
%We do this for two reasons: one, it is easier to trade Lagrangian parameters for $\MAonesq$ and $\MAtwosq$ than the CP-even eigenmasses, and two, there is a near-degeneracy among the doublet eigenstates (CP-even, CP-odd and charged Higgses, with mass $\MAonesq$) and among the singlet eigenstates (CP-even and CP-odd, with mass $\MAtwosq$).
Using these approximations the one-loop effective potential due to the heavy Higgs scalars now evaluates to
%%%%%%%%%%%%%%%%%%%%%%%%%%%%%%%%
\bea
\Delta V \propto  \sum_{i} \left[a_i  M^4_{A,i}  + 2 b_{i}M^2_{A,i}  +  (b_i^2  + 2 c_i)\right] \left( \log\frac{M^2_{A,i}}{Q^2} - \frac{3}{2}\right)
\label{eq:effpotperturb}
\eea
%%%%%%%%%%%%%%%%%%%%%%%%%%%%%%%
where $a_i$ are constants and the field-dependent coefficients $b_i$ and $c_i$ are obtained from the hatted coefficients in Eq.~(\ref{eq:fielddependentmass}).
Reducing $\Delta V$ to this form considerably simplifies the calculation of Higgs mass corrections.
$\Delta V$ as presented here must also ensure that the decoupling behavior discussed in the previous subsection is preserved at one-loop order.
This result is demonstrated in Appendix~\ref{apx:decoupling}.

The full details of our computation and the corresponding results are presented in Appendix~\ref{apx:effpotderiv},
where two cases satisfying the condition $\left(M^2_A\right)_{12} = 0$ in Eq.~(\ref{eq:massmatcpodd}) were considered.
In the first case, which we call Case (A), we assume that the one-loop corrections arise from a single heavy scale $M_A = M_{A_D} = M_{A_S}$.
The results from this case will be used in our discussion of phenomenology in Section~\ref{sec:pheno}.
In the second case, which we call Case (B), we show the effect of splitting the CP-odd Higgs masses, thereby obtaining corrections from two heavy scales. In this case we set the terms $\Al, \mu', A_\kappa, m_3, m'_S$ to zero for simplicity.
Further, we ignored corrections that depend on electroweak couplings since we are
interested in the limit $\lambda \gg g$.
It is important to note that Cases (A) and (B) pertain not only to different
limits of the mass spectra of the CP-odd scalars,
but also to somewhat different regions of the Lagrangian parameters.
In Case (A), the parameters  $\Al, \mu', A_\kappa, m_3, m'_S$ can be non-zero in general,
with the condition $\left(M^2_A\right)_{12} = 0$ imposing $\Al = \mu'$.
On the other hand, Case (B) explicitly sets them all to zero.

% In the basis of Eq.~(\ref{eq:bogdanbasis}), the correction to the Higgs mass is obtained as \textbf{(AM: Maybe we should give the % correction to each element in the gauge basis? Also equality is not correct in the below equation only true decoupling!)}

For Case B, the one-loop self-energy corrections obtained in the basis ($h^0_u,h^0_d,h^0_s$) are
%%%%%%%%%%%%%
%%%%%%%%%%%%%%%%
\bea
\nonumber \Pi_{11} &=& \frac{\lambda^4 v^2}{16\pi^2} s^2_\beta \left[ - (4 c_{2\beta}+c_{4\beta}+1) \log\left(\frac{\MAonesq}{M^2_Z}\right) +  2 \log\left(\frac{\MAtwosq}{M^2_Z}\right)  \right], \nonumber \\
\nonumber \Pi_{12} &=&  \frac{2\lambda^4 v^2}{16\pi^2} s_{\beta} c_{\beta}  (2+c_{4\beta}) \log\left(\frac{\MAonesq}{M^2_Z}\right), \nonumber \\
  \nonumber \Pi_{22} &=& \frac{\lambda^4 v^2}{16\pi^2} c^2_\beta \left[ -  (-4 c_{2\beta}+c_{4\beta}+1) \log\left(\frac{\MAonesq}{M^2_Z}\right)
  + 2 \log\left(\frac{\MAtwosq}{M^2_Z}\right)  \right], \nonumber \\
   \label{eq:fullcorrec_mainsec}
 \nonumber \Pi_{13} &=& \frac{\lambda^3 v \mue}{16\pi^2} s_{\beta} \left[ -(1+3 c_{2\beta}) \log\left(\frac{\MAonesq}{M^2_Z}\right)
  + 4  \log\left(\frac{\MAtwosq}{M^2_Z}\right) \right],
  \\
  \nonumber \Pi_{23} &=& \frac{\lambda^3 v \mue}{16\pi^2} c_\beta \left[  -(1-3 c_{2\beta}) \log\left(\frac{\MAonesq}{M^2_Z}\right)
  +  4 \log\left(\frac{\MAtwosq}{M^2_Z}\right) \right],
   \\
  \nonumber \Pi_{33} &=& \frac{4 \lambda^2 \mue^2}{16\pi^2}  \log\left(\frac{\MAonesq}{M^2_Z}\right).
\eea
%%%%%%%%%%%%%%%
When these contributions are rotated into the basis of Eq.~(\ref{eq:bogdanbasis}), we get the
(1, 1) element of the self-energy corrections as
%%%%%%%%%%%%%%%%%%%%%%%%%%%%%%%%
\bea
\Pi_{hh} = \frac{\lambda^4 v^2 s_\beta}{16\pi^2} \left[ \left(c^2_\beta(2+c_{4\beta}) - s^2_\beta(1+c_{4\beta}+4c_{2\beta})\right) \log\left(\frac{\MAonesq}{M^2_Z}\right) + 2s^2_\beta \log\left(\frac{\MAtwosq}{M^2_Z}\right)  \right]. \nulein
\label{eq:finalcorrec_Bogdan}
\eea
%%%%%%%%%%%%%%%%%%%%%%%%%%%%%%%%%
 This is a good approximation for the Higgs mass correction
 when the mixing between the $SU(2)$ Higgs doublets and the singlet is negligible.
At large $\tan \beta$, Eq.~(\ref{eq:finalcorrec_Bogdan}) further simplifies to
%%%%%%%%%%%%%%%%%%%%%%%%%%%%%%%%
\bea
\Pi_{hh} \xrightarrow{\text{large} \tan\beta} \frac{\lambda^4 v^2}{16\pi^2} \left[ 2 \log\left(\frac{\MAonesq}{M^2_Z}\right) + 2 \log\left(\frac{\MAtwosq}{M^2_Z}\right)  \right].
\label{eq:finalcorrec_Bogdan_largeTB}
\eea
%%%%%%%%%%%%%%%%%%%%%%%%%%%%%%%%%

%%%%%%%%%%%%%%%%%%%%%%%%%%%%%%
\begin{figure}
\begin{center}
\includegraphics[width=5.9cm]{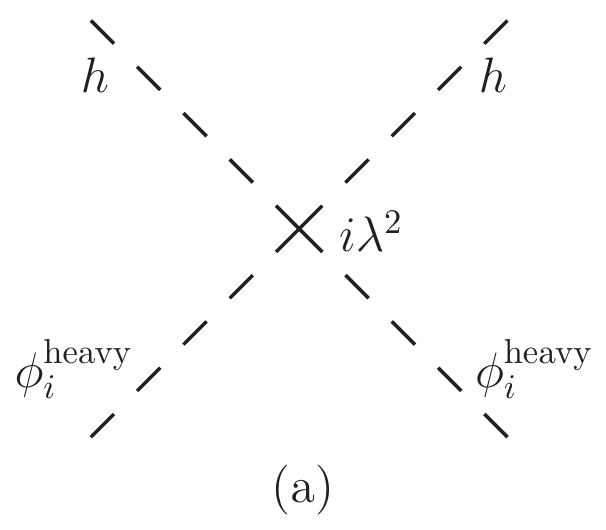}
\quad \quad \quad
\includegraphics[width=5.1cm]{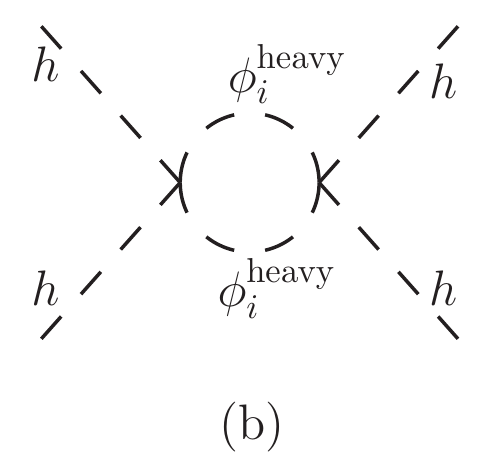}
\caption{\footnotesize (a): Tree level quartic vertices involving at least two $h$ fields with vertex factors $\propto \lambda^2$,
in the limit $\tan\beta \gg 1$.
In this limit, $h^0_u \ra h, h^0_d \ra H, h^0_s \ra h^0_s$.
No $h^4$ quartic terms at formed at tree level.
$\phi_i$ correspond to the heavy fields $H, h^0_s, A^0_D, A^0_S$.
(b): One-loop quartic vertices with four $h$ legs, formed from the tree level vertices in (a).
These are $\propto \lambda^4$ and account for most of the radiative corrections to the Higgs mass in our model.}
\label{fig:quartics}
\end{center}
\end{figure}
%%%%%%%%%%%%%%%%%%%%%%%%%%%%%%

  We could gain an intuitive understanding of our results
   by qualitatively estimating the size of the one-loop radiative
 corrections without recourse to the effective potential.
 Such an estimate would serve as a useful cross-check of the results obtained from $\Delta V$.
 We do this by the following argument in our limit of interest, $\tan \beta \gg 1$ and $\lambda \gg g$.
 In this limit, we identify
 the real scalars $h^0_u \ra h,~h^0_d \ra H,~h^0_s \ra h^0_s$,
 where $h$ is the SM Higgs boson, and $H$ and $h^0_s$ are non-standard Higgses.
 The Standard Model Higgs and the Goldstone bosons reside mostly in $H_u$
 and the non-standard CP-even and CP-odd Higgses in $H_d$ and $S$.
 % It is then useful to rewrite Eq.~(\ref{eq:higgsafterEWSB}) as
 %%%%%%%%%%%%%%%%
% \bea
% H_u &\xrightarrow{\tan \beta \gg 1}& \left( \begin{array}{c}
 %   G^+      \\
%    v + \frac{1}{\sqrt{2}} (h + i G^0)
 %      \end{array} \right),
% \nonumber \\
% H_d &\xrightarrow{\tan \beta \gg 1}& \left( \begin{array}{c}
 %     \frac{1}{\sqrt{2}} (H + i A_D^0)   \\
 %      H^-
 %      \end{array} \right).
% \nonumber \\
% S &\xrightarrow{\tan \beta \gg 1}&  \frac{1}{\sqrt{2}}(s+ h^0_s + i A^0_S),
% \label{eq:higgsafterEWSB_beta90}
 % \eea
 %%%%%%%%%%%%%%%%

 For $\lambda \gg g$, the most important quartic terms at tree-level are those proportional to $\lambda^2$.
 Before EWSB, we can read them off from Eq.~(\ref{eq:higgspot_allterms})
 as the terms
 $|H_u^0|^2 |H_d^0|^2$,
 $H_u^0 H_d^0 H^+_u H^-_d$,
 $|H_u^0|^2|S|^2$
 and $|H_d^0|^2|S|^2$.
 After EWSB, we
 can expand $H_u, H_d, S$ using Eq.~(\ref{eq:higgsafterEWSB}) to obtain various quartic vertices in terms of the real and charged scalars.

Fig.~\ref{fig:quartics}(a) shows all the tree-level quartic vertices that involve at least two $h$ fields.
Recall that the SM Higgs mass is set by the coupling strength of the quartic term $h^4$ in the scalar potential.
 The tree-level $\lambda$-dependent quartic $h^4$ terms are suppressed at large $\tan \beta$.
 However, using the vertices in Fig.~\ref{fig:quartics}(a), we can construct four one-loop level quartic vertices proportional to $h^4$, as shown in Fig.~\ref{fig:quartics}(b).
Each of these diagrams is proportional to $\lambda^4 \log (M^2_{A_i}/M^2_Z)$, where $M^2_{A_i}$
is the mass scale of the heavy field running in the loop.
Two diagrams each correspond to $\MAonesq$ and $\MAtwosq$ respectively.
Since the internal propagators are identical, each diagram comes with a factor of 2. Canonical normalization of the mass term of a real scalar implies an additional factor of $1/2$. Finally, including the loop factor $1/16\pi^2$,  we find the approximate correction to the lightest CP-even eigenstate to be
 %%%%%%%%%%%%%%%%%%%%%%%%%%%%%%%%
 \bea
 \Pi_{hh} \approx \frac{1}{2} \cdot \frac{2 \cdot 2 \cdot \lambda^4}{16\pi^2} \left[\log \left( \frac{\MAonesq}{M^2_Z} \right)  + \log \left( \frac{\MAtwosq}{M^2_Z} \right) \right],
\label{eq:loopestimate_differentscales}
 \eea
 %%%%%%%%%%%%%%%%%%%%%%%%%%%%%%%%
 which agrees with Eq.~(\ref{eq:finalcorrec_Bogdan_largeTB}).

It would be interesting to compare the Higgs mass corrections obtained from the heavy Higgs fields
and those obtained from top squarks.
For simplicity, let us set the pseudoscalar masses equal, $M_A = M_{A_D} = M_{A_S}$, and obtain
 %%%%%%%%%%%%%%%%%%%%%%%%%%%%%%%%
 \bea
 \Pi^{\rm higgs}_{hh} = \frac{\lambda^4}{4\pi^2} v^2 \log \left( \frac{M^2_A}{M^2_Z} \right) .
 \label{eq:loopestimate_higgses}
 \eea
 %%%%%%%%%%%%%%%%%%%%%%%%%%%%%%%%
Again for simplicity, we can assume the top squarks are degenerate ($m_{\tilde{t}} = m_{\tilde{t_1}} = m_{\tilde{t_2}}$). Then we obtain \cite{Ellwanger:2009dp}
  %%%%%%%%%%%%%%%%%%%%%%%%%%%%%%%%
 \bea
 \Pi^{\rm stops}_{hh} = \frac{3 y^4_t}{4\pi^2} v^2 \log \left( \frac{m^2_{\tilde{t}}}{M^2_Z} \right)  .
 \label{eq:loopestimate_stops}
 \eea
 %%%%%%%%%%%%%%%%%%%%%%%%%%%%%%%%
The factor of 3 arises from the three QCD colors.
 If the pseudoscalars and the top squark are degenerate ($M_A = m_{\tilde t}$), we find
 from Eqs.~(\ref{eq:loopestimate_higgses}) and (\ref{eq:loopestimate_stops}) that $\Pi^{\rm higgs}_{hh} \gsim \Pi^{\rm stops}_{hh}$ for $\lambda \gsim 3^{1/4} y_t$.
 Since $y_t = m_t/v \simeq 1$,
we have $\Pi^{\rm higgs}_{hh} \gsim \Pi^{\rm stops}_{hh}$ for $\lambda \gsim 1.3$.

In the discussion of our model's phenomenology, we set $m_{\tilde{t}} = 800~\gev$
 while $M_A$ ranges between 4~TeV and 8~TeV; therefore,
 the one-loop corrections from the Higgs sector dominate those from the stops.
 Hence, throughout our analysis, the effect of the top squark correction to the SM Higgs mass is neglected.

%%%%%%%%%%%%%%%%%%%%%%%%%%%%%%%%%%%%%%%%%%%%
\begin{figure}[t]
\center
\includegraphics[width=0.48\textwidth]{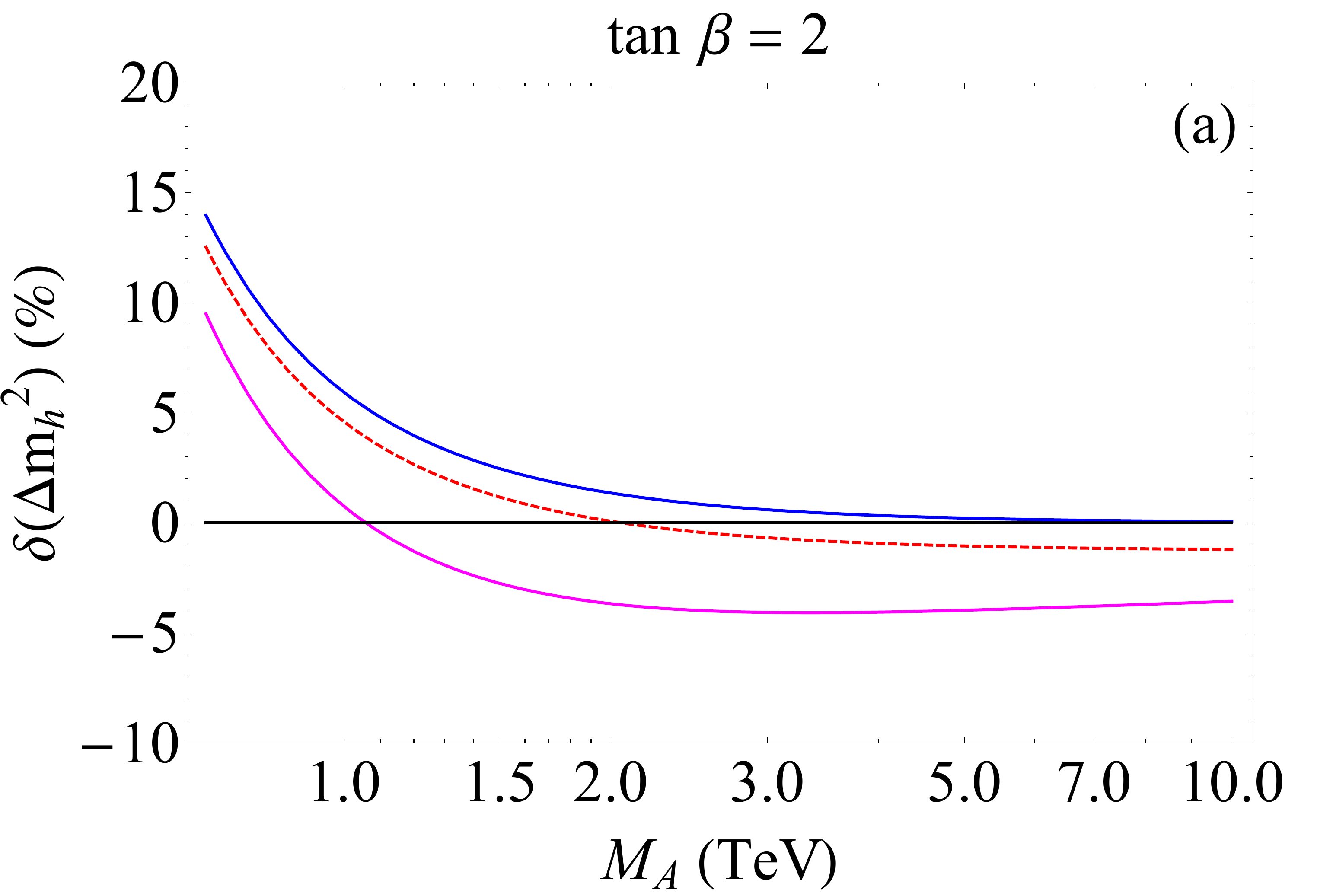}
\includegraphics[width=0.48\textwidth]{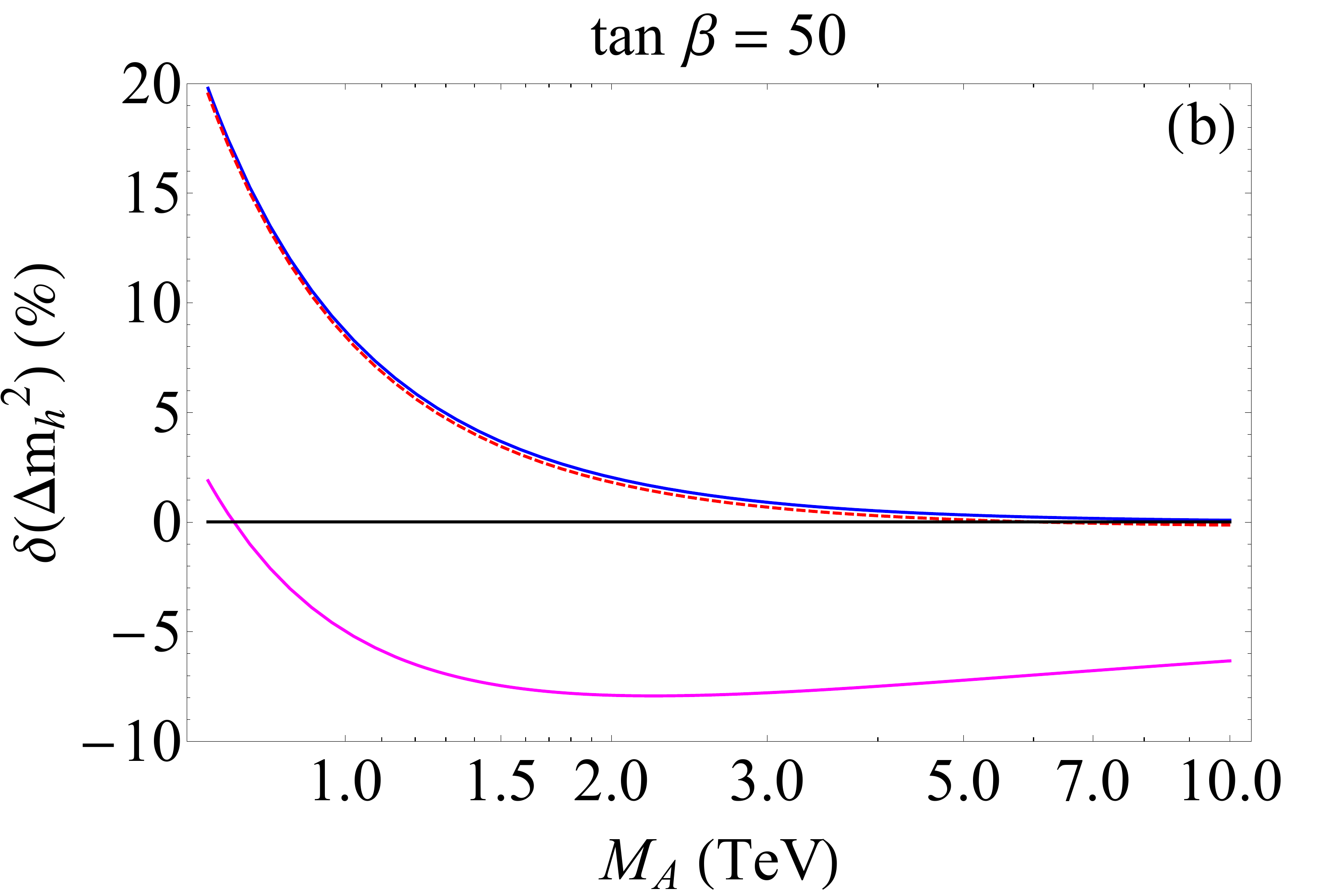}
\caption{\footnotesize Discrepancies between the Higss mass radiative corrections obtained from our one-loop effective
potential in Eq.~(\ref{eq:fullcorrec_mainsec}) and those obtained by other means, as a function of the mass $M_A$ of degenerate pseudoscalars.
The blue, dashed red and magenta curves represent corrections obtained from Eq.~(\ref{eq:finalcorrec_Bogdan}),
Eq.~(\ref{eq:loopestimate_differentscales}) and Ref.~\cite{Ellwanger:2005fh} respectively.
(a) corresponds to $\tan \beta = 2$, (b) corresponds to $\tan \beta = 50$.
See text for details of the behavior of the curves.}
\label{fig:4levelcorrexns}
\end{figure}
%%%%%%%%%%%%%%%%%%%%%%%%%%%%%%%%%%%%%%%%%%%

We can now quantify the discrepancies between the results obtained by a full one-loop effective potential calculation and those obtained by other means.
To do so, first we compute the correction to the Higgs squared mass
obtained from Eq.~(\ref{eq:fullcorrec_mainsec}), and denote it by $\Delta m^2_h$.
For the same set of parameters, we compute $(\Delta m^2_h)_i$ for each alternative approximation labelled by $i$.
We then take the difference and normalize it to $\Delta m^2_h$ and define the discrepancy as
%%%%%%%%%%%%%%%%
\bea
\delta(\Delta m^2_h) = {(\Delta m^2_h)_i - \Delta m^2_h \over \Delta m^2_h},
\eea
%%%%%%%%%%%%%%%%%
which is then expressed as a percentage.
This approach eliminates the $\lambda$-dependence of the discrepancies and allows us to focus on their behavior with respect
to $\tan \beta$ and the heavy (pseudo)scalar masses.

Assuming for simplicity that the CP-odd scalars are degenerate, we depict in Fig.~\ref{fig:4levelcorrexns}
 the discrepancies as a function of $M_A$.
 Figs.~\ref{fig:4levelcorrexns}(a) and \ref{fig:4levelcorrexns}(b) correspond to $\tan \beta = 2$ and
 $\tan \beta = 50$ respectively.
 The blue curve denotes $(\Delta m^2_h)_i$ obtained from the approximation in
Eq.~(\ref{eq:finalcorrec_Bogdan}).
Since this approximation neglects doublet-singlet mixing, it tends to
overestimate the correction, i.e., $\delta(\Delta m^2_h) > 0$ as observed in the plot.
The discrepancy is also seen to asymptote to zero at large $M_A$, where the CP-even singlet Higgs decouples from the SM Higgs.
The dashed red curve is  $(\Delta m^2_h)_i$ obtained from our qualitative diagrammatic estimate (Eq.~(\ref{eq:finalcorrec_Bogdan_largeTB})).
Since the estimate is designed for large $\tan \beta$ it disagrees with the blue curve at $\tan \beta = 2$,
but coincides with it very well at $\tan \beta = 50$.
The magenta curve depicts $(\Delta m^2_h)_i$ obtained from
NMSSMTools 4.5.1~\cite{Ellwanger:2005fh,Ellwanger:2009dp},
which also computes the one-loop radiative corrections from
the effective potential, albeit under a different set of approximations.
We find an interesting discrepancy here, to which we now turn.

\begin{figure}
 \begin{center}
 \includegraphics[width=0.48\textwidth]{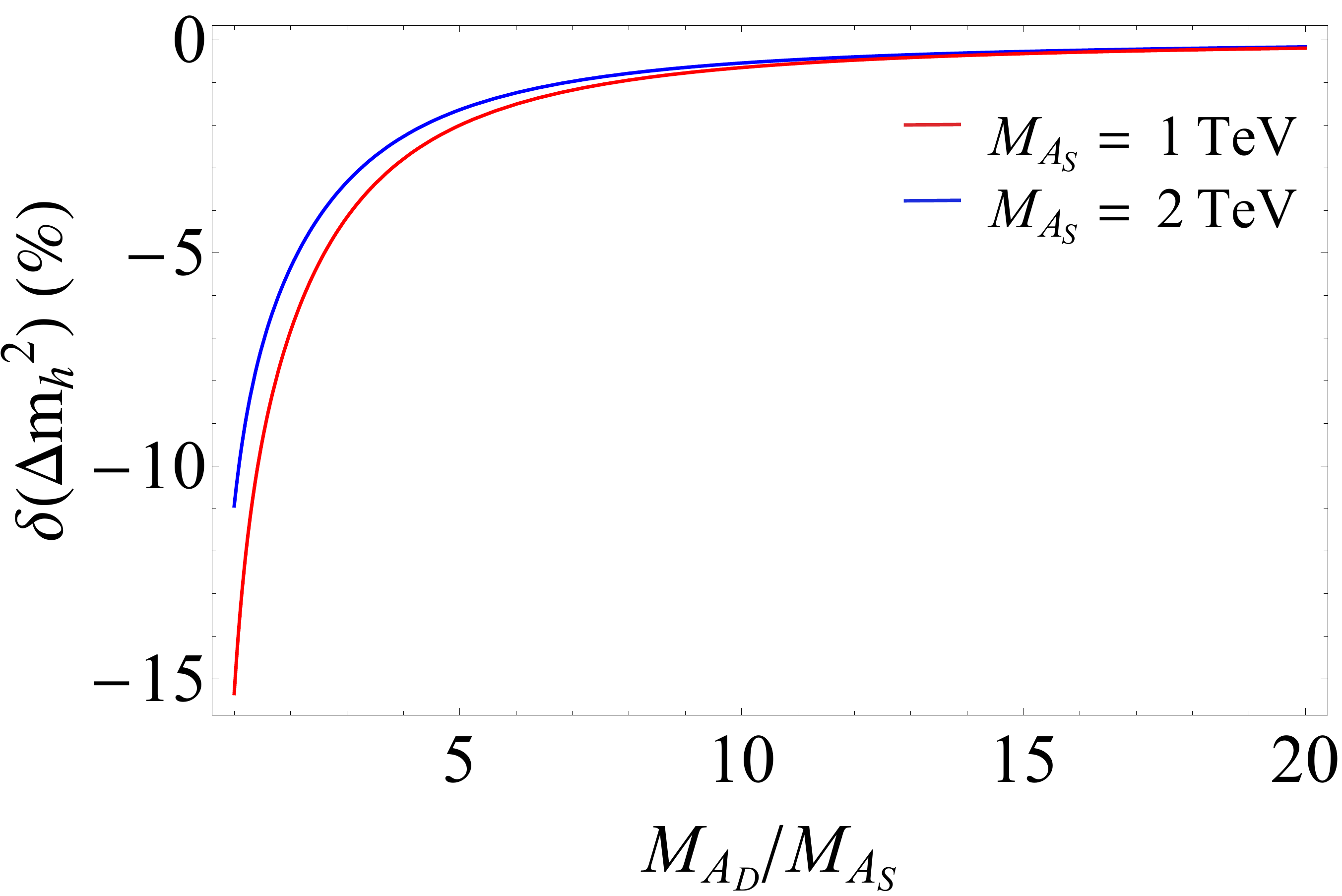}
 \caption{\footnotesize The discrepancy between Higgs mass corrections obtained by Ref.~\cite{Ellwanger:2005fh} (which
 were used in the original code of NMSSMTools 4.5.1) and by us, as
 a function of the ratio of the heavy CP-odd Higgs masses.
 The red (blue) curve corresponds to $M_{A_S} = 1 (2)~\tev$.
 The discrepancy arises due to an approximation assumed by Ref.~\cite{Ellwanger:2005fh}, namely, that a
 hierarchy exists in the pseudoscalar spectrum.
 It is seen that our results agree when there is indeed a hierarchy.
 See text for more details.}
 \label{fig:discrep_tools_3}
 \end{center}
 \end{figure}

The eigenvalues of the CP-odd mass matrix in Eq.~(\ref{eq:massmatcpodd}) are given by
%%%%%%%%%%%%%%%
\bea
E^2_{\pm} = {1 \over 2}\left(T \pm \sqrt{T^2 - 4 D} \right),
\eea
%%%%%%%%%%%%%%%
where $T = \left(M^2_A\right)_{11} + \left(M^2_A\right)_{22}$ is the trace and
$D = \left(M^2_A\right)_{11}\left(M^2_A\right)_{22} - \left(M^2_A\right)^2_{12}$ is
the determinant of the mass matrix.
In Ref.~\cite{Ellwanger:2005fh}, it is assumed that $D \ll T^2$, so that the eigenmasses are
obtained as $E^2_+ \simeq T,~E^2_- \simeq D/T$.
This always leads to a hierarchy between the pseudoscalar masses.
In contrast, our approach sets the off-diagonal element $\left(M^2_A\right)_{12}$ to zero so that the eigenmasses
are $E^2_+ = \MAonesq = \left(M^2_A\right)_{11},~E^2_- = \MAtwosq =\left(M^2_A\right)_{22}$.
Therefore, our approach allows for a variety of mass splittings.
Hence the discrepancy between us and Ref.~\cite{Ellwanger:2005fh} is
expected to be maximum when the CP-odd Higgses are
degenerate, and minimum when these masses are well split.
We illustrate this effect in Fig.~\ref{fig:discrep_tools_3}.
Since $\left(M^2_A\right)_{12} = 0$ in our approach,
we set $\left(M^2_A\right)_{12}$ to zero in the expression of Ref.~\cite{Ellwanger:2005fh}
as well, in order to make an ``apples-to-apples" comparison.
We then plot $\delta(\Delta m^2_h)$
as a function of $M_{A_D}/M_{A_S}$, where we have taken $\lambda = 1.25$, $\tan \beta = 50$ and $\mue = 110~\gev$.
The red and blue curves depict $M_{A_S} = 1~\tev$ and $M_{A_S}=2~\tev$ respectively.
As expected, we find the discrepancy at its greatest at $M_{A_D}/M_{A_S} = 1$, which
can reach upto $\sim 15\%$.
Observe also that $\delta(\Delta m^2_h) < 0$, implying that
Ref.~\cite{Ellwanger:2005fh} underestimates the one-loop contribution to the Higgs mass
in the region around $M_{A_D}/M_{A_S} = 1$.
As we raise $M_{A_D}/M_{A_S}$, the discrepancy drops quickly and our results concur.

The results of Ref.~\cite{Ellwanger:2005fh} were originally used in the code of NMSSMTools 4.5.1 \cite{Ellwanger:2009dp}.
Since our phenomenology in Section~\ref{sec:pheno} assumes $M_{A_D} = M_{A_S}$,
we replaced the code in NMSSMTools 4.5.1 with the expressions that we derived in Appendix~\ref{apx:effpotderiv}.

 %%%%%%%%%%%%%%%%%%%%%%%%%%%%%%
%%%%%%%%%%%%%%%%%%%%%%%%%%%%%%

\subsubsection{Stability of the electroweak scale}

The minimization conditions of the tree level Higgs potential in Eq.~(\ref{eq:Higgspottree})
lead to the same relation between the electroweak scale and the SUSY parameters seen in the MSSM.
In particular, the EWSB condition is~\cite{Martin:1997ns}
%%%%%%%%%%%%%%%%%%%%%%%%%%%%
\bea
M_Z^2 = \frac{t_\beta^2 +1}{t_\beta^2 - 1} \left(m_{H_d}^2 - m_{H_u}^2\right) - \left(m_{H_u}^2 + m_{H_d}^2 \right)- 2 \left| \mue \right|^2,
\eea
%%%%%%%%%%%%%%%%%%%%%%%%%%%%
which at large $\tan \beta$ reduces to
%%%%%%%%%%%%%%%%%%%%%%%%%%%%
\bea
\frac{1}{2}M_Z^2 \approx -  m_{H_u}^2 -  \left|\mue \right|^2,
\label{eq:ewsb_cond}
\eea
%%%%%%%%%%%%%%%%%%%%%%%%%%%
where the $m_{H_d}^2$ terms are suppressed by $t_\beta^{-1}$.
With this result we can now quantify the relative importance of different contributions
(denoted by $a$) to the EWSB scale ($M_Z^2/2$) as
%%%%%%%%%%%%%%%%%%%%%%%
\bea
\Delta (a^2) = \left| \frac{a^2}{M_Z^2/2}\right|.
\eea
%%%%%%%%%%%%%%%%%%%%%%%
 The tree-level and one-loop corrections are the same as in the MSSM and are well-known~\cite{Dimopoulos:1995mi}.
 For instance, the tree-level contribution due to $\mue \lsim 350$~GeV is equivalent to the one-loop contribution of stops at $m_{\tilde{t}} \lsim 800$~GeV~\cite{Kribs:2013lua}.
 Hence the regions we are considering in this article are typically as tuned as regions of the MSSM with a light stop.

%%%%%%%%%%%%%%%%%%%%%%%
\subsubsection{Higgs couplings to SM particles}

LHC measurements of signal strengths (production rate $\times$ branching ratio) can
potentially constrain the properties of the Higgs sector.
Mixing among the Higgs fields can in principle alter the lightest Higgs boson's SM-like behavior.
We follow the analysis of Ref.~\cite{Barbieri:2013hxa} to apply the relevant limits.

After including the one-loop self-energy corrections, we rotate the Higgs fields $(h^0_u, h^0_d, h^0_s)$ into the mass eigenbasis
$(h_1, h_2, h_3)$ and identify the lightest scalar as
%%%%%%%%%%%%%%%%
\bea
h_1 = ( -h^0_u~\sin \alpha + h^0_d~\cos \alpha )\cos \gamma+ h^0_s~\sin \gamma,
\label{eq:higgs_admixture}
\eea
%%%%%%%%%%%%%%%%
where the angles $\alpha$ is the usual MSSM CP-even mixing angle that characterizes doublet-doublet mixing
and $\gamma$ characterizes the doublet-singlet mixing.
We can then write down the reduced couplings of $h_1$ to pairs of fermions and vector bosons as
%%%%%%%%%%%%%%%%
\bea
{g_{tth_1} \over g_{tth_{\rm SM}}} &=& \cos \gamma \left( \cos \delta + {\sin \delta \over \tan\beta}  \right), \nulein
{g_{bbh_1} \over g_{bbh_{\rm SM}}} &=& \cos \gamma ( \cos \delta - \sin \delta \tan \beta ), \nulein
 {g_{VVh_1} \over g_{VVh_{\rm SM}}} &=& \cos \gamma \cos \delta,
\eea
%%%%%%%%%%%%%%%%
where $\delta = \alpha - \beta + \pi/2$.

 If we inspect the off-diagonal entries of Eq.~(\ref{eq:decoupling_tree_Bogdan}),
 we see that for  $\Al \ll M_A$ and large $\tan \beta$,
 $\left(M^2_H\right)_{hH} < \left(M^2_H\right)_{hS}$.
 Thus as we raise $M_A$,
 the heavy doublet Higgs (identified as $h_3$) generally decouples faster than the heavy singlet
 (identified as $h_2$),
 % viz.,
% $\alpha$ goes to zero much faster than $\gamma$,
as noted by Refs.~\cite{Hall:2011aa,Farina:2013fsa}
In dealing with the phenomenological consequences of our model, we focus exactly on the region of $\Al \ll M_A$ and large $\tan\beta$.
Therefore for the rest of this analysis we assume $h_3$ is decoupled from the
spectrum and $h_2$ is not.
In this limit, the mixing angle $\gamma$ is given by

%%%%%%%%%%%%%%%%%%
\bea
\sin^2 \gamma = {m^2_{hh} - m^2_{h_1} \over m^2_{h_2} - m^2_{h_1}},
\eea
%%%%%%%%%%%%%%%%%%
where $m^2_{hh} = \lambda^2 v^2 \sin^22\beta + M^2_Z \cos^22\beta$, and the Higgs couplings to fermions
and vector bosons become
%%%%%%%%%%%%%%%%%
\bea
{g_{tth_1} \over g_{tth_{\rm SM}}} = {g_{bbh_1} \over g_{bbh_{\rm SM}}} = {g_{VVh_1} \over g_{VVh_{\rm SM}}} = \cos \gamma.
\eea
%%%%%%%%%%%%%%%%%%

Using these relations Ref.~\cite{Barbieri:2013hxa} performed a
universal fit on the LHC signal strength measurements and found that $\sin^2 \gamma \leq 0.23$ at 95\% C.L.
This result was obtained using tree level relations for the reduced couplings.
When we include our one-loop corrections, we find that the reduced couplings are modified by less than $1\%$.
Therefore, in the discussion of our model's phenomenology in Section~\ref{sec:pheno}
we will simply use the results of Ref.~\cite{Barbieri:2013hxa} to constrain the Higgs couplings with LHC measurements.

%SSSSSSSSSSSSSSSSSSSSSSSSSSSSSSSSSSSSSSSSSS
\subsection{Neutralino sector}
\label{subsec:neut_sector}

The composition of the lightest neutralino and its couplings to the Higgs sector is
central to the dark matter phenomenology of our model.
The neutralino mass matrix in the basis ($\bino, \wino, \hdino, \huino, \sino$) is given by
%%%%%%%%%%%%%%%%%%%%
\begin{equation}
\Mneut =
\left( \begin{array}{ccccc}
M_1  &  0    & - g_1 v \cos\beta/\sqrt{2} &  g_1 v \sin \beta/\sqrt{2}  & 0 \\
0  & M_2 & g_2 v \cos \beta/\sqrt{2} & -g_2 v \sin \beta/\sqrt{2}  & 0 \\
- g_1 v \cos\beta/\sqrt{2}   &  g_2 v \cos \beta/\sqrt{2}  & 0  & -\mue & - \lambda v \sin\beta  \\
    g_1 v \sin \beta/\sqrt{2}  & -g_2 v \sin \beta/\sqrt{2}   &  -\mue  &   0  &  - \lambda v \cos \beta \\
        0   &  0  &  - \lambda v \sin\beta  &  - \lambda v \cos \beta   &  \mu' \\
\end{array} \right)
\label{eq:massmat_neutralinos}
\end{equation}
%%%%%%%%%%%%%%%%%%%%%
Notice that when $\mu' \ll M_1, M_2, \mue$, large $\lambda$ couplings imply a large
Higgsino component in the lightest neutralino.
This feature has many unique consequences for the dark matter phenomenology discussed in Sec.~\ref{sec:pheno}.
As we shall see, the Higgs-$\none$-$\none$ coupling strengh plays an important role in
constraining our model with dark matter experiments.
This coupling, denoted hereafter by $g_{h\chi\chi}$, is obtained as
%%%%%%%%%%%%%%%%%%%%%%%
\bea
g_{h\chi\chi} =
\frac{\lambda}{\sqrt{2}}  (\zeta_{H_u} N_{\hdino} N_{\sino} + \zeta_{H_d} N_{\huino} N_{\sino} + \zeta_{S} N_{\huino} N_{\hdino})
- \frac{g_1}{2} N_{\bino} (\zeta_{H_u} N_{\hdino} - \zeta_{H_d} N_{\huino}), \nonumber \\
 \label{eq:Couphfull}
 \eea
%%%%%%%%%%%%%%%%%%%%%%
where the $N_i$ and $\zeta_j$ are the appropriate components of the lightest neutralino
and the SM-like Higgs respectively.
In terms of the rotation angles in Eq.~(\ref{eq:higgs_admixture}), we can read off
%%%%%%%%%%%%%%%%%%
\bea
\zeta_{H_u} = -\sin \alpha \cos \gamma ,~~~~\zeta_{H_d} = \cos \alpha \cos \gamma ,~~~~\zeta_{S} = \sin \gamma .\nonumber
\eea
%%%%%%%%%%%%%%%%%

The dominant channel for $\none$-nucleon scattering is through a $t$-channel Higgs.
Therefore, dark matter direct detection experiments, as well as
limits on the invisible decay width of the Higgs,
apply strong contraints on the coupling $g_{h\chi\chi}$.
A suppressed $g_{h\chi\chi}$ can occur in our model either when
the Higgsino content is suppressed, making $\none$ mostly singlino or bino, or when there is
a delicate cancellation between the various terms in Eq.~\ref{eq:Couphfull}.
We illustrate this point in more detail in Sec.~\ref{sec:pheno}.

%%%%%%%%%%%%%%%%%%%%%%%%%%%%%%%%%%%%%%%%%%%%
\begin{figure}[t]
\center
\includegraphics[width=0.4\textwidth]{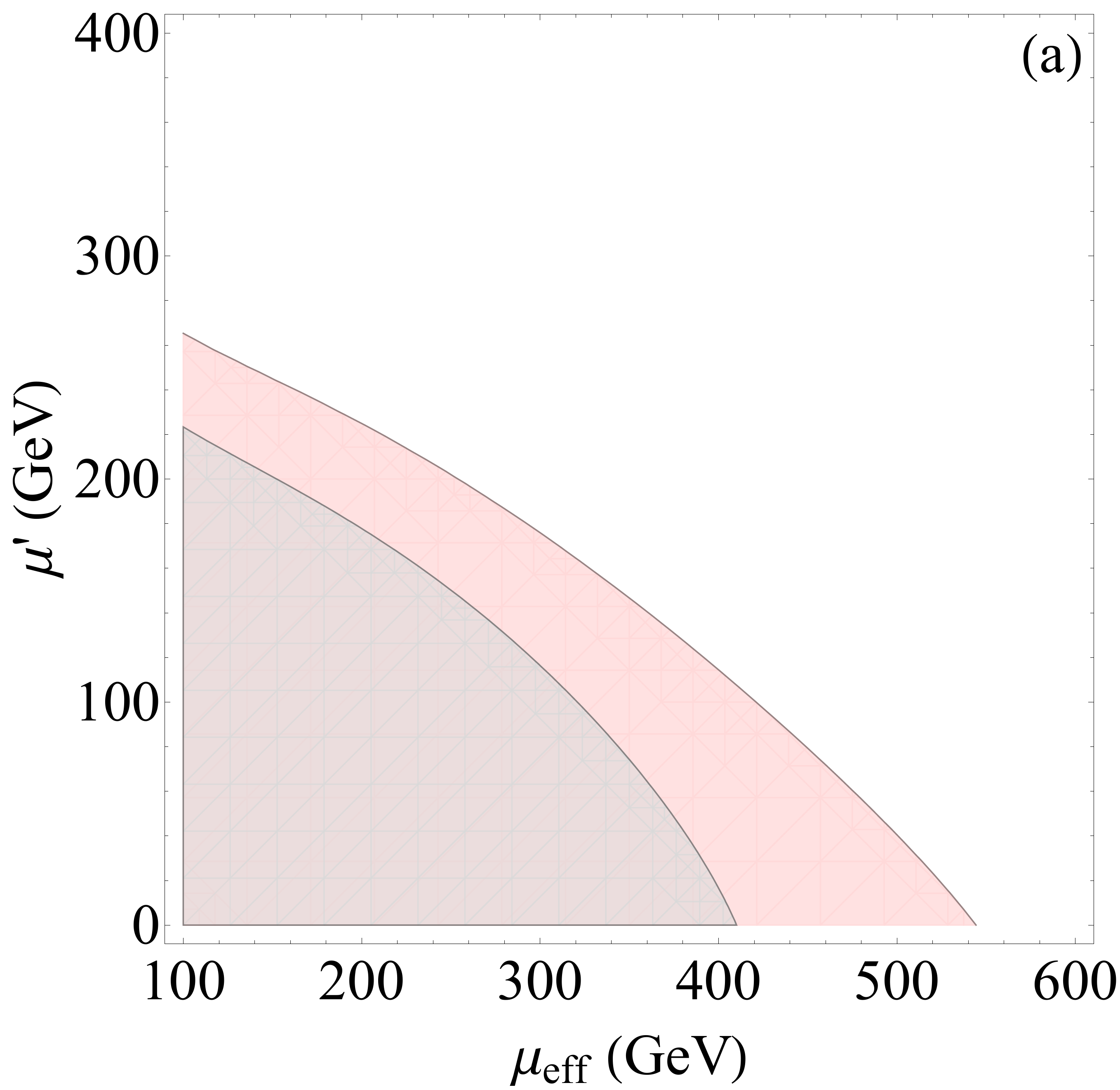}
\quad \quad
\includegraphics[width=0.39\textwidth]{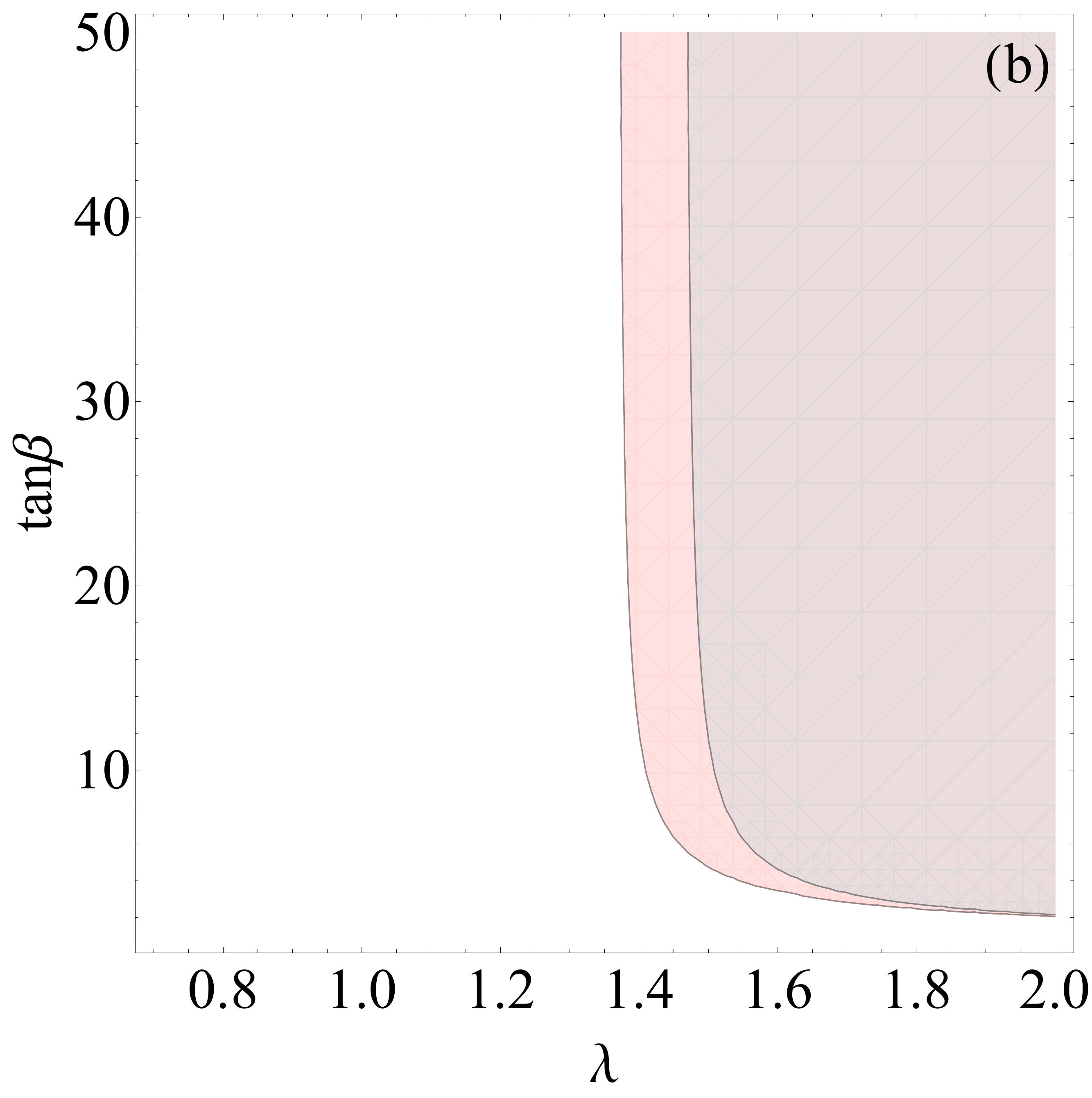}
\caption{\footnotesize Limits from electroweak precision parameter $T$ on the neutralino sector of our model.
The shaded regions are where $\Tneut > 0.15$ and therefore excluded at $95 \%$ C.L.
Regions shaded gray correspond to the wino decoupled from the
spectrum ($\Mwino = 10~\tev$) and regions shaded red to $\Mwino = 200~\gev$.
In (a), $\lambda = 1.25$ and $\tan\beta = 5$
and in (b), $\mue = \mu' = 300~\gev$.
See text for details of the behavior of these curves.}
\label{fig:ewpt}
\end{figure}
%%%%%%%%%%%%%%%%%%%%%%%%%%%%%%%%%%%%%%%%%%%
\subsubsection{Electroweak precision limits}
\label{subsec:ewpt_limits}

Due to mixing between the Higgsinos and the singlino induced by large $\lambda$ in
certain regions,
constraints from electroweak precision experiments can be strong in Fat Higgs/$\lambda$-SUSY models ~\cite{Barbieri:2006bg, Franceschini:2010qz}.
In particular, the $T$ parameter can get large contributions from the neutralino sector, denoted hereafter by $\Tneut$.
This phenomenon is understood easily in the limit where the electroweak gauginos $\bino$ and $\wino$ decouple from the spectrum, i.e., $M_1, M_2$ are very large.
This leaves us with three mass scales $\mue, \mu'$ and $\lambda v$,
which set the mass of the lightest neutralino, $\Mchione$.
The lightest chargino is mostly Higgsino with a mass $\mue$.
In this limit, $\Tneut$ is large when $\Mchionepm-\Mchione$ is large and when there is as a significant Higgsino component in $\none$.
For simplicity, let us work in the limit where $\tan \beta$ is large.
Then the neutralino mass matrix in Eq.~(\ref{eq:massmat_neutralinos}) is simply
%%%%%%%%%%%%%%%%%%%%
\begin{equation}
\Mneut \sim
\left( \begin{array}{ccc}
0  & -\mue & - \lambda v  \\
-\mue  &   0  &  0 \\
- \lambda v  &  0   &  \mu' \\
\end{array} \right).
\label{eq:massmatneut}
\end{equation}
%%%%%%%%%%%%%%%%%%%%%
$\Tneut$ is suppressed either when $\mu' \sim \mue \sim \lambda v$, where
$\Mchionepm-\Mchione$ is small, or when $\mue \gg \mu' \sim \lambda v$,
 where the Higgsino component in $\none$ is
suppressed.
For $\mu' \ll \mue \sim \lambda v$, where both $\Mchionepm-\Mchione$
  and the Higgsino component in $\none$ are large, constraints from $\Tneut$
can be strong.

 Lowering the mass of the wino triplet $M_2$ to $\sim \mue \sim \lambda v$
 can have a significant impact on $\Tneut$.
This is because the wino would mix with the light neutralinos and charginos.
Lowering the bino mass $M_1$, on the other hand, gives only a negligible contribution to $\Tneut$.
 This is because the bino mixing with the rest of the neutralinos is only proportional to $g_1$.\footnote{
It must be remembered that relative minus signs between $\mue, \mu'$ and $\Mwino$ would introduce quantitative changes in the picture owing to new phases in the neutralino mixing angles.
We will not include these relative signs in our discussion.}

% For large $\tan \beta$ and small $\mu'$, the spectrum reduces to a heavy Dirac fermion with mass $\sqrt{\lambda^2 v^2 +
%\mu^2_{\rm eff}}$ and a lighter Majorana fermion, $\none$, with mass  $\sim (\lambda^2 v^2\mu_{\rm eff} s_{2\beta})/(\lambda^2 v^2 + \mu^2)$.
% If the lightest neutralino has a large Higgsino component, $\Tneut$ receives a large contribution due to a large splitting between $\none$ and the chargino with a mass $\sim \mu$.
% Increasing $\mu_{\rm eff}$ reduces this effect because the $\none$ becomes more singlino-like, while increasing the magnitude of $\mu'$ also reduces $\Tneut$ as the splittings between the chargino and the neutralinos become typically smaller.

In Fig.~\ref{fig:ewpt} we present the $T$-parameter contributions from the charginos and neutralinos,
which were computed using the general expressions provided in Ref.~\cite{Martin:2004id}.
In Fig.~\ref{fig:ewpt}(a), we take $\lambda = 1.25$ and $\tan \beta = 5$ and show our results in the
$\mue-\mu'$ plane.
The shaded regions denote where $\Tneut$ is not within the 95\% C.L range $[-.01,0.15]$ set by the Particle Data Group~\cite{Agashe:2014kda}.
The gray region corresponds to large gaugino masses $(M_1,M_2) = (10\mbox{ TeV},10\mbox{ TeV})$ while the red region corresponds to a light wino with $(M_1,M_2) = (10\mbox{ TeV},0.2\mbox{ TeV})$.
As discussed above, lowering the wino mass can lead to a larger $\Tneut$.
For small $\mu'$, $\Tneut$ decreases as $\mue$ increases due to a reduction in the Higgsino component of the lightest neutralino.
Similarly, raising $\mu'$ has the effect of reducing the splittings between the neutralinos and charginos which also leads to a smaller $\Tneut$.

The effects of varying $\lambda$ and $\tan \beta$ on $\Tneut$ are presented in Fig.~\ref{fig:ewpt}(b).
Here we fix $\mue = \mu' = 300$~GeV.
The colored regions have the same definition as those in Fig.~\ref{fig:ewpt}(a).
Since the elements of $\Mneut$ quickly asymptote to fixed values as a function of $\tan \beta$, it can be seen that $\Tneut$ is insensitive to large $\tan \beta$. This insensitivity to large $\tan \beta$ is clear in the relation derived in Ref.~\cite{Barbieri:2006bg}
%%%%%%%%%%%%%%%%%%%%%%%%%%%%%%
\bea
T_\chi \approx \left(\frac{t_\beta^2-1}{t_\beta^2+1}\right)^2 F(\mu_{\rm eff},\mu',\lambda),
\eea
%%%%%%%%%%%%%%%%%%%%%%%%%%%%%%%
where $F(\mu_{\rm eff},\mu',\lambda)$ is some function of these variables. This relation also shows that $T_\chi$ is suppressed as $t_\beta$ approaches 1, thereby
allowing for larger values of $\lambda$.
%As we lower $\tan \beta$, the Higgsino-singlino mixing in Eq.~(\ref{eq:massmat_neutralinos}) decreases and leads to a larger allowed region. This mixing increases with $\lambda$, which is why $\Tneut$ increases as we scan from left to right.
As stated before, lowering $M_2$ typically increases the neutralino and chargino contributions to the $T$-parameter.
 However, it is important to emphasize that increasing either $\mue$ or $\mu'$ can significantly lower the electroweak precision constraints even for large $\tan \beta$. A large $\mue$ comes at the cost of a slight increase in electroweak fine-tuning, but
 can greatly weaken $T$-parameter constraints.

Finally, we make two remarks.
First, the $S$-parameter was not discussed here.
This is because the contributions of our model to $S$ are very small in our regions of interest
and hence the constraints are much weaker than those on the $T$-parameter.
Second, the $T$-parameter receives a stop-sbottom contribution,
 as discussed in Ref.~\cite{Barbieri:2006bg}.
In the limit of zero left-right mixing, this is given by
%%%%%%%%%%%%%%%%%
\bea
T_{\rm st-sb} \approx 0.05 \left( {500~\gev \over m_{\tilde t_L}} \right)^2
\eea
%%%%%%%%%%%%%%%%%
In our phenomenological discussions, we will choose $m_{\tilde t_L} = 800~\gev$ to suppress this contribution.

%Using the general expressions for neutralino and chargino contributions to the $T$ parameter in Ref.~\cite{Martin:2004id},
% we present the effects of varying parameters $\mu_{\rm eff}, \mu', M_1, M_2, \tan \beta$ and $\lambda$ in Fig. ~\ref{Fig:ewpt}. % As the present Particle Data Group (PDG) limits allow for $T$ to be in the range $ 0.07 \pm 0.08$ at 95\% C.L.~\cite{Agashe:2014kda}, the lines in Fig.~\ref{Fig:ewpt} correspond to $\Tneut = 0.15$ where the contributions are smaller for larger values of $%\mu_{\rm eff}$ and $|\mu'|$. The left and right panel correspond to $(\lambda, \tan \beta = (2,2)$ and $(1,50)$ respectively. The solid lines corresponds to a decoupled scenario where $M_1 = M_2 = 10$~TeV and  agrees with Fig.3 of Ref.~\cite{Barbieri:2013hxa}. The dashed line shows the effect of lowering the Bino mass ($M_1 = 200$~GeV) while the dotted line shows the effect of lowering the Wino mass  ($M_2 = 200$~GeV). The dot-dashed line shows the effect of lowering both gaugino masses to be $M_1 = M_2 = 200$~GeV.  The case of $(\lambda, \tan \beta = (2,50)$ is highly constrained while most of the regions in $\mu_{\rm eff}-\mu'$ space are allowed when $(\lambda, \tan \beta) = (1,2)$.

\section{Phenomenology}
\label{sec:pheno}

In this section we study the phenomenological constraints on the large
 $\tan \beta$ regions of the Fat Higgs/$\lambda$-SUSY models.
In addition to the constraints arising from Higgs corrections discussed in the previous section,
we also include limits from dark matter experiments, most importantly those set by the LUX experiment~\cite{Akerib:2013tjd}.
In particular, the mass and couplings of the lightest neutralino $\none$ can put strong
constraints on our parameter space.

In order find phenomenologically viable regions, we modified NMSSMTools 4.5.1~\cite{Ellwanger:2009dp}
to include the Higgs mass corrections we computed in Sec.~\ref{subsec:radiatifcorrex}.
 We then made the following simplifying assumptions:
%%%%%%%%%%%%%%
\begin{itemize}
\item In the Higgs sector, we take the pseudoscalars to be degenerate, with $M_{A_D} = M_{A_S} = M_A$.
     Furthermore we assume that $m'_S = m_3 = 0$, so that the heavy CP-even Higgs bosons are also (nearly) degenerate. The condition that the CP-odd masses are degenerate requires that $\left(M^2_A \right)_{12} = 0$ in Eq.~\ref{eq:massmatcpodd}, which implies $A_\lambda = \mu'$.
     Therefore, both $\mu'$ and $\mue$ control the amount of doublet-singlet mixing in Eq.~(\ref{eq:massmatcpeven}).
     The only independent parameters in the Higgs sector are then: $\lambda, \mue, \mu', \tan \beta$ and $M_A$.
\item In order to be safe from electroweak precision bounds, we decouple the winos at $M_2 = 10$~TeV, leading to an effective theory for the neutralino system with five free parameters : $M_1$, $\mue$, $\mu'$, $\lambda$ and $\tan\beta$.
\item We require $\mue > 104$ GeV to evade the LEP II bound on charged Higgsinos \cite{LEPbound}.
\item The sleptons and the first two generations of squarks are decoupled from the low energy phenomenology and their masses set at 5 TeV, unless stated otherwise.
    The top squark parameters are set at $m_{\tilde{Q}_3} = m_{\tilde{U}_3}=  800$~GeV and $A_t=0$,
    thereby making the stop contributions to the Higgs mass and the electroweak symmetry breaking condition in Eq.~(\ref{eq:ewsb_cond}) small.
    This choice of stop masses also avoids constraints from collider searches~\cite{atlas_stops_semi,atlas_stops_had,cms_stops} and, as mentioned in
    Sec.~\ref{subsec:ewpt_limits}, from electroweak precision tests.
\item  We choose to require the conventional upper limit $\tan \beta \leq 60$, so that $y_b \leq 1$ at the weak scale. Larger values of $y_b$ may be allowed as long they do not develop a Landau pole at a scale below that of $\lambda$.
\item We assume that the relic density of dark matter is the value determined by Planck \cite{Ade:2015xua}.
Hence, in scenarios where $\Omega_\chi h^2 < 0.12$, we assume some nonthermal mechanism for generating the observed
relic abundance.
\end{itemize}
%%%%%%%%%%%%%%

These assumptions reduce the number of independent SUSY parameters to
%%%%%%%%%%%%%%
\bea
\lambda,~\tan \beta,~M_A,~\mue,~\mu',~M_1~. \nonumber
\eea
%%%%%%%%%%%%%%%
As discussed in Sec.~\ref{subsec:ewpt_limits}, precision electroweak
constraints are weak either when $\mu'$ or $\mu_{\rm eff}$ are large for any
$\tan \beta$, or when $\mu' \ll \lambda v \sim \mue$ at low $\tan \beta$. In
these regions, $g_{h\chi\chi}$ (as defined in Eq.~\ref{eq:Couphfull}) can also be
found to satisfy dark matter direct detection and relic density constraints.
In particular we find the following viable parametric regions.

%(a) {\bf Singlino Dark Matter: Large $\tan \beta$ and $\mu' \lsim \lambda v \sim \mue \gg M_1$}.\\
%$\none$ is a mostly singlino state with a small enough Higgsino component to avoid dark matter constaints. In order to generate the observed cosmological relic abundance we consider either resonant annihilation through the Higgs or co-annihilation with a nearly degenerate sfermion. This scenario is a uniquely NMSSM scenario.

%(b) {\bf Bino Dark Matter: Large $\tan \beta$ and $M_1 \lsim \lambda v \sim \mue \gg \mu'$}.\\
%$\none$ is a mostly bino state with a small enough Higgsino component to avoid dark matter constaints. This scenario is the bino equivalent of the previous scenario. This region of parameter space is similar to that of large $\tan \beta$ in the MSSM.

%(c) {\bf The Well-tempered Neutralino: Low $\tan \beta$ and $\mu' \ll \lambda v \sim \mue \sim M_1$}\\ $\none$ is an admixture of bino, Higgsino and singlino. As $\mu'$ is small the mass of $\none$ is typically suppressed. Therefore a very delicate cancellation between the various components is needed to find a viable region~\cite{ArkaniHamed:2006mb}.

%In each of these regions, constraints from the Higgs boson mass, direct detection experiments, the dark matter relic abundance and the invisible decay width of the $Z$ and Higgs boson are crucial in determining viable parametric scenarios.

\subsection{Singlino DM: Large $\tan \beta$ and $\mu' < \lambda v \sim \mue \ll M_1$}
Large $\lambda$ and large $\tan \beta$ are a new region of parameter space that
have not been emphasized in the literature before. We showed in
Sec.~\ref{subsec:higgs_sector} that this region can be compatible with the mass
of the SM Higgs boson because one-loop radiative corrections to the Higgs mass
are insensitive to $\tan\beta$ at large values, and are set solely by $\lambda$
and $M_A$. We also showed that precision electroweak constraints can be weak in
this region. We now show that this region is also compatible with constraints
from dark matter.

As mentioned in Sec.~\ref{subsec:neut_sector}, it can be seen from
Eq.~(\ref{eq:Couphfull}) that $g_{h\chi\chi}$ is suppressed when $\none$ is
mostly singlino such that $N^2_{\sino} \simeq 1$. This requirement is possible
when $\mu'$ is relatively small compared to the other mass scales in the
neutralino mass matrix. The annihilation of $\none$ into SM fields in the early
universe is generally inefficient, due to both the $Z$- and $h$-mediated
channels being suppressed by the small Higgsino component of $\none$.
Therefore, for the
cosmological relic abundance to be below the observed value $\Omega_\chi h^2
\simeq 0.12$, we consider the mechanism of resonant annihilation and
co-annihilation~\cite{Griest:1990kh}.

\subsubsection{Resonant annihilation}
\begin{figure}
\begin{center}
\includegraphics[width=0.44\textwidth]{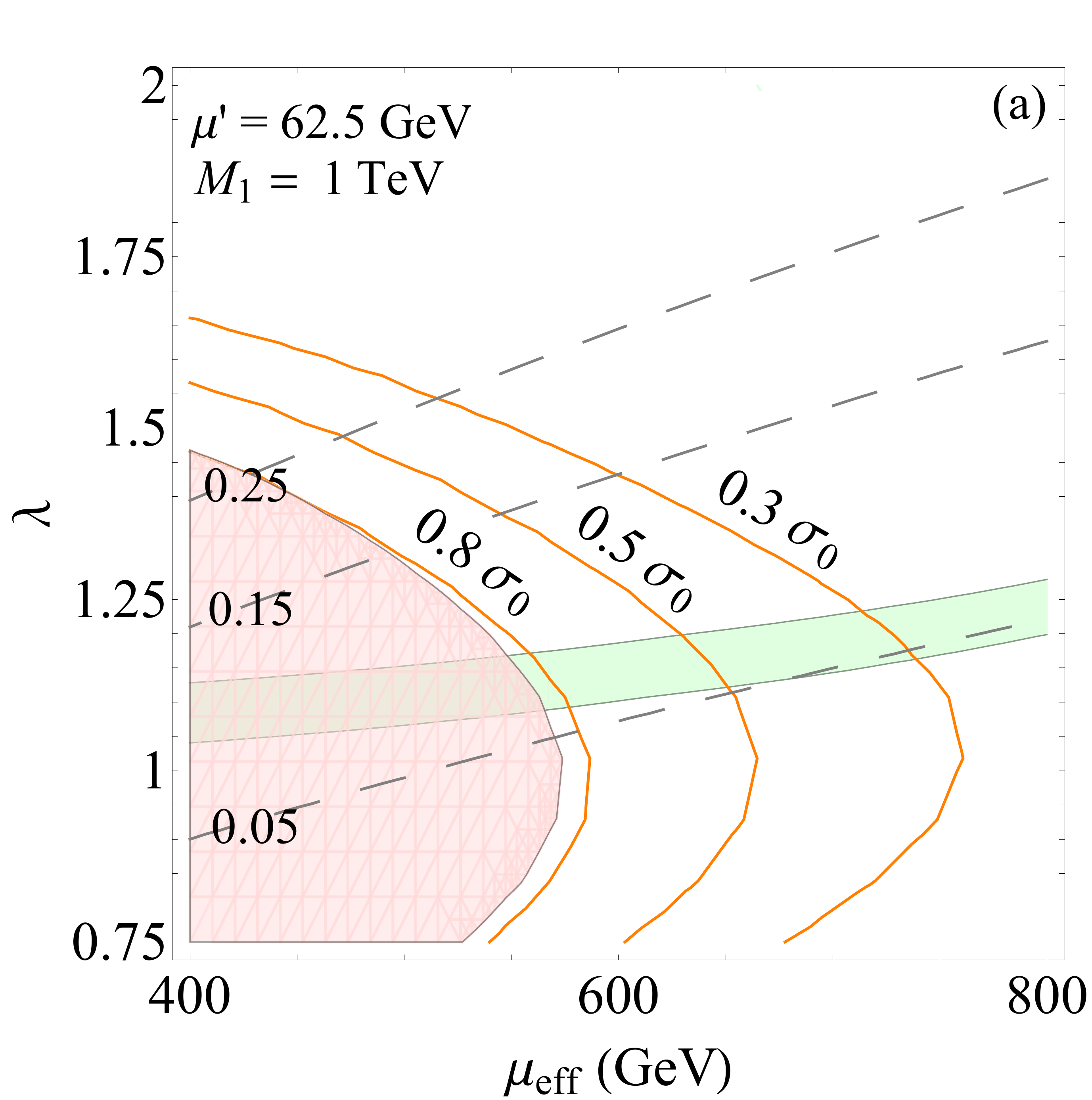}
\includegraphics[width=0.45\textwidth]{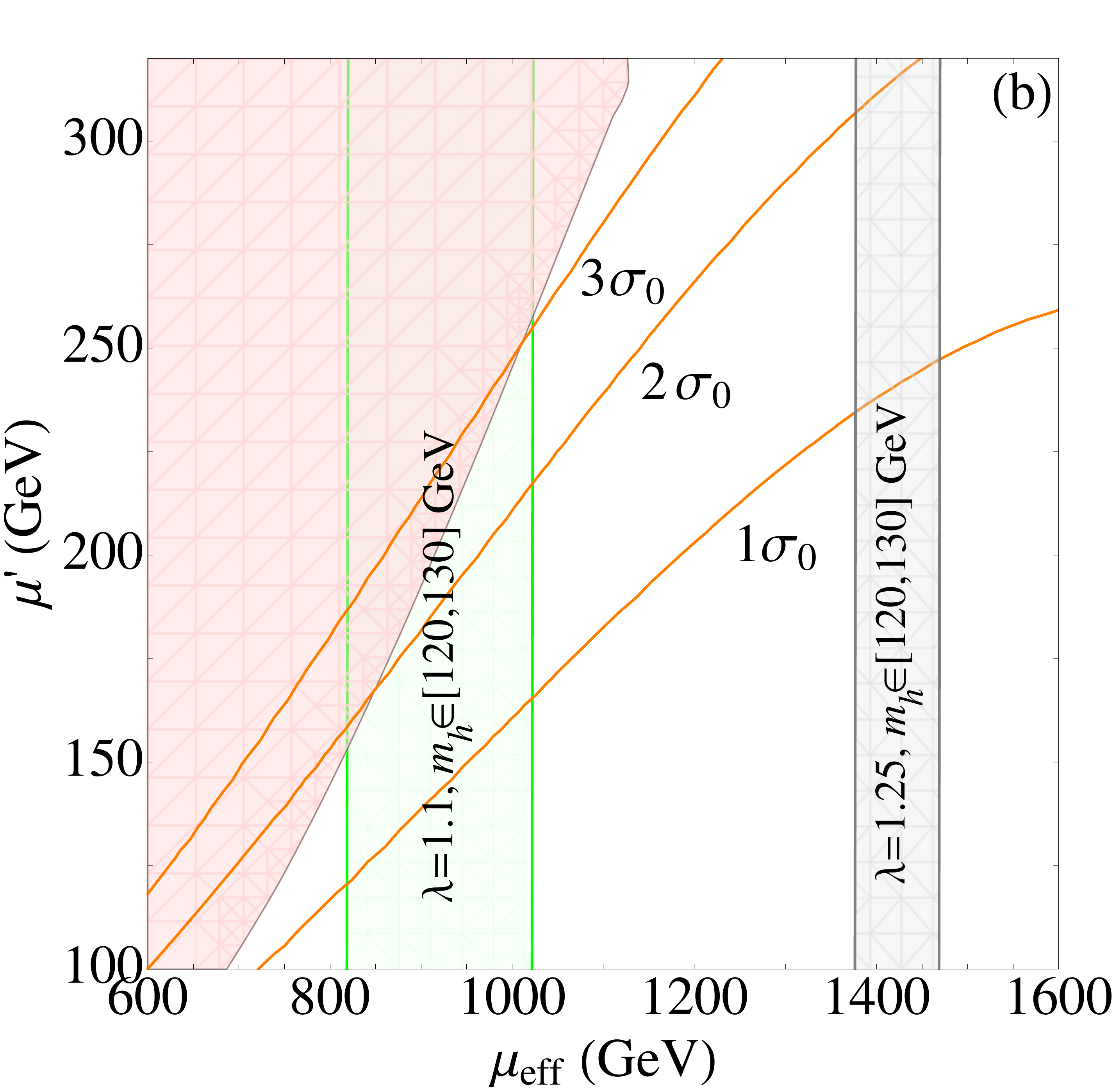} \\
\includegraphics[width=0.44\textwidth]{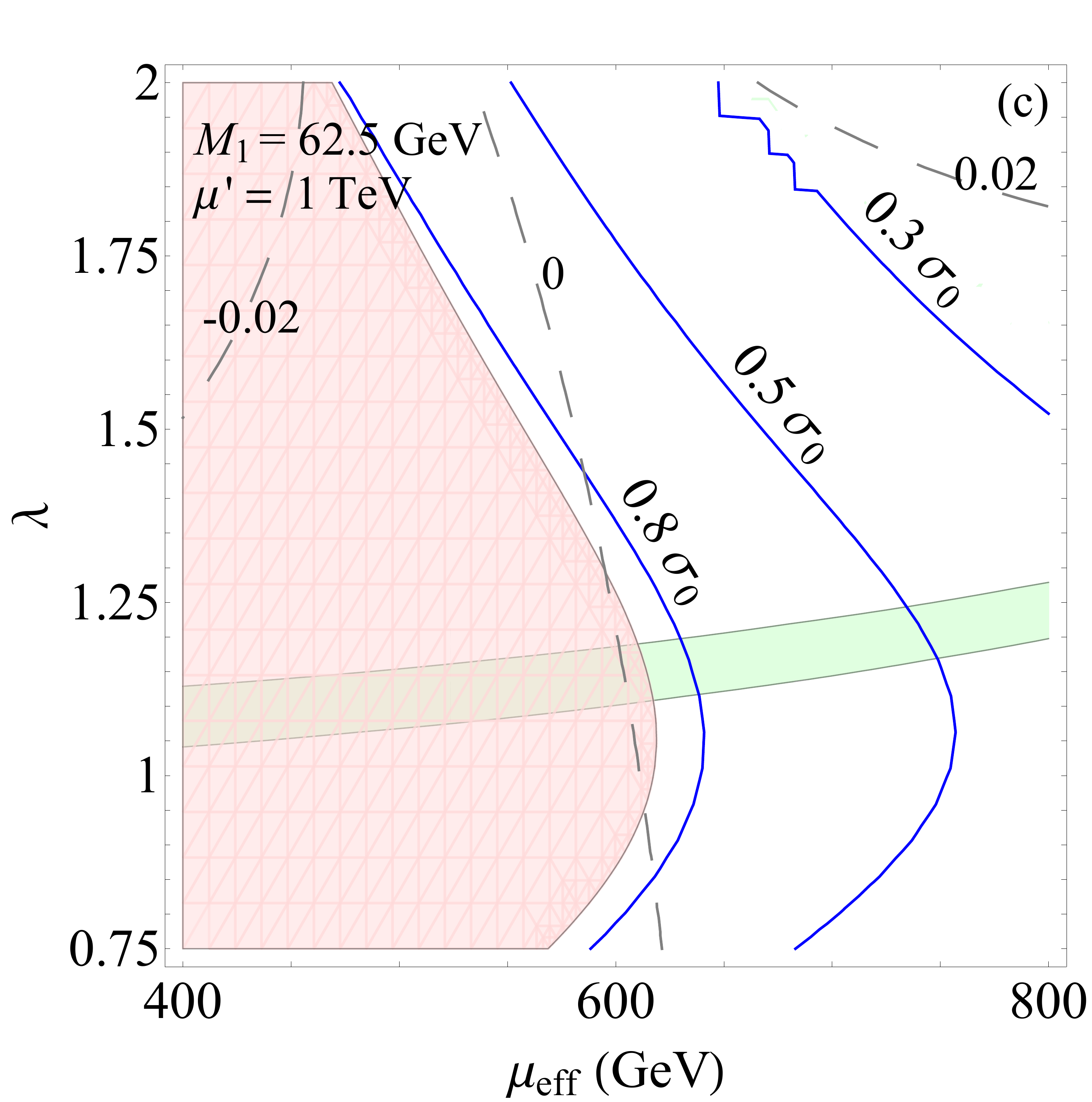}
\includegraphics[width=0.45\textwidth]{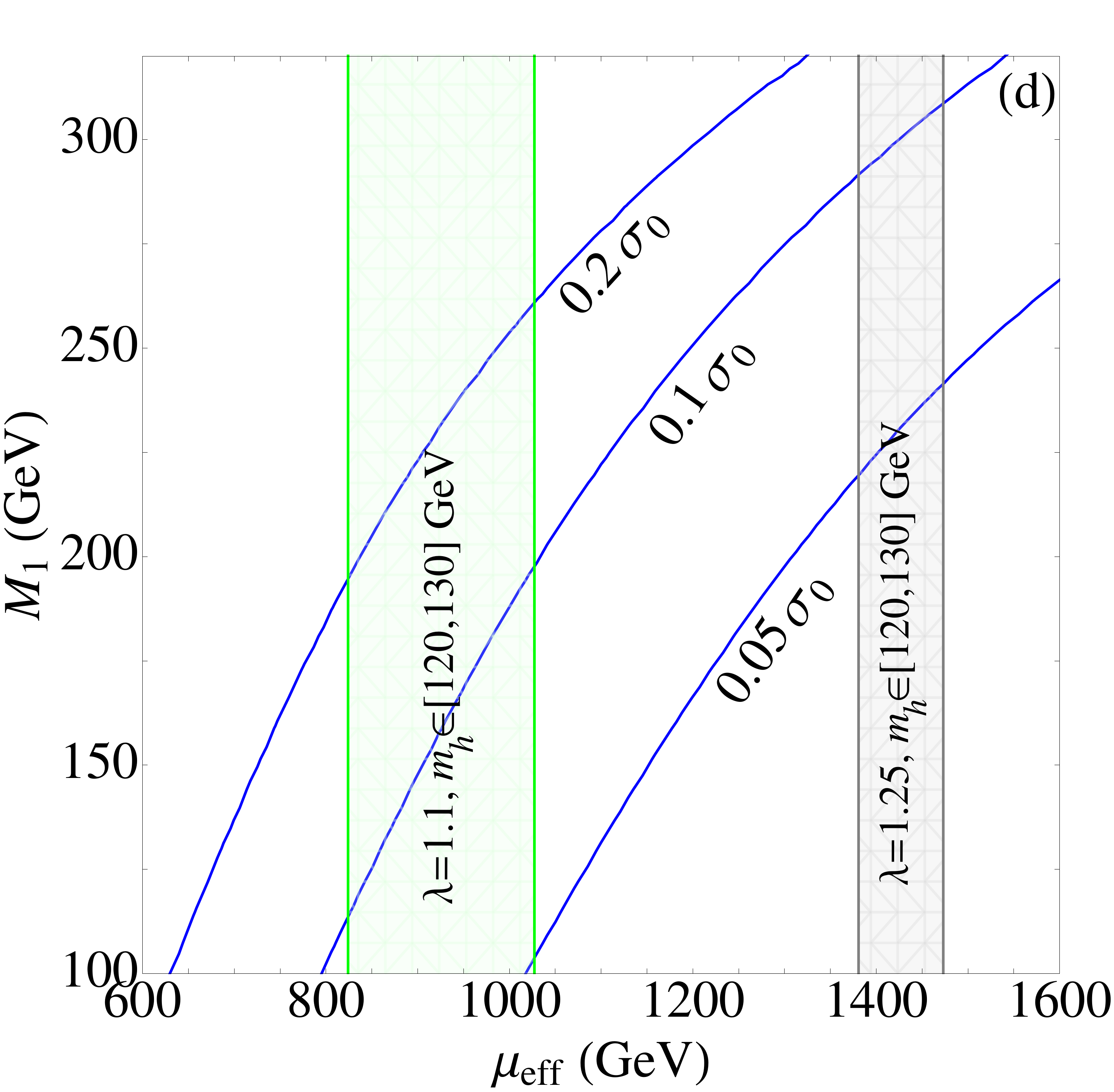}
\caption{\footnotesize Large $\tan \beta$ parametric scenarios for
Fat Higgs/$\lambda$ SUSY models.
(a) Singlino resonant annihilation ($\mu' =
62.5$~GeV, $M_1 = 1$~TeV): orange curves correspond to $\sigma_{\rm SI}$ in units
of $\sigma_0 = 10^{-45}~\text{cm}^2$.
The red shaded region is excluded by LUX at 90\% C.L.
and the green shaded region corresponds to $120~\gev < m_h < 130 ~\gev$.
The dashed lines are contours of $\Tneut$. (b) Singlino co-annihilation: orange
curves and red region the same as (a). The green and gray shaded regions correspond to
$120~\gev < m_h < 130 ~\gev$ for  $\lambda = 1.1$ and $\lambda = 1.25$
respectively. (c) Bino resonant annihilation ($\mu' = 1$~TeV, $M_1 = 62.5
$~GeV): blue curves correspond to $\sigma_{\rm SI}$, with the remaining colors the
same as in (a). (d) Bino co-annihilation: blue curves correspond to $\sigma_{\rm SI}
$, with the remaining colors remaining the same as in (b).
In all these plots we have set $\tan \beta = 50$. The critical features are explained in the text.
}
\label{fig:bisi}
\end{center}
\end{figure}

If $\Mchione$ happens to be close to $m_h/2$, it can undergo resonant
annihilation through an $s$-channel Higgs.
Therefore, we set $\mu' = 62.5$~GeV in this scenario.
We also set $\mue = 800$~GeV, $M_1 = 1$~TeV, $\tan\beta = 50$ and
$M_A = 4$~TeV.
The orange curves in Fig.~\ref{fig:bisi}(a) depict
contours of the LSP-nucleon scattering rates, $\sigma_{\text{SI}}$ (in units of
$\sigma_0 = 10^{-45}~\text{cm}^2$), on the $\mue-\lambda$ plane.
The red shaded regions are excluded by LUX at
90\%~C.L., and the green band corresponds to $120~ \text{GeV} \leq m_h \leq 130~ \text{GeV}$.
Contours of $\Tneut$ are denoted by dashed curves.

In Fig.~\ref{fig:bisi}(a), the dark matter-nucleon scattering rates are smaller
for larger $\mue$ because the Higgsino fraction in $\none$ decreases.
$\Tneut$ is observed to rise with increasing $\lambda$ due to an increase in
the Higgsino
fraction of $\none$. The region around $m_h \sim 125$~GeV corresponds to
$\Tneut \sim 0.05$, which is safe from electroweak precision constraints.
This regions is also safe from invisible Higgs decay bounds since the process
$h \rightarrow \none \none$ is phase space suppressed.

%%%%%%%%%%%%%%%%%%%%%%%%%%%%%%
%%%%%%%%%%%%%%%%%%%%%%%%%%%%%%

\subsubsection{Co-annihilation region}
If the mass spectrum is such that one or more sfermions are nearly degenerate
with $\none$, dark matter annihilation could be assisted by the sfermions
through co-annihilation effects, leading to a small relic abundance. Bounds
from LEP on charged sfermions~\cite{LEPsleptonbounds} would then imply that
$\Mchione \gsim 104$ GeV.

We investigate this scenario in Fig.~\ref{fig:bisi}(b). In this figure we
assume that the correct thermal dark matter relic abundance is generated by a
process like co-annihilation. We do not explicitly state the mass spectrum
or compute the resultant relic abundance. Again the orange curves in
Fig.~\ref{fig:bisi}(b) are contours of $\sigma_{\text{SI}}$ (in units of $\sigma_0
= 10^{-45}~\text{cm}^2$) on the $\mue-\mu'$ plane. We have chosen $\lambda = 1.25
$, $\tan \beta = 50$, $M_1 = 1$ TeV and $M_A = 8$ TeV. The larger values
of $\mu_{\rm eff}$ compared to those in  Fig.~\ref{fig:bisi}(a) imply a greater
amount doublet-singlet mixing. Therefore a slightly larger value of $M_A$ is
chosen in this scenario as compared to that of  Fig.~\ref{fig:bisi}(a). The
region shaded red is excluded by LUX at 90$\%$ C.L. The green (gray) vertical
bands correspond to $m_h \in [120,130]$ GeV for $\lambda = 1.1$ ($1.25$).
 The effect of varying $\lambda$ on $\sigma_{\text{SI}}$ is not shown
since the scattering cross-section is insensitive to it due to the large values
of $\mue$. The decrease in $\sigma_{\text{SI}}$ with $\mue$ is due
to the decoupling of the Higgsinos leading to the suppression of $g_{h\chi\chi}$.
The increase in $\sigma_{\text{SI}}$ with $\mu'$ is due to the larger Higgsino
fraction in $\none$, which leads to an enhanced $g_{h\chi\chi}$. The relatively
large size of $\mue$ and $\mu'$ here suppress the Higgsino sector contributions
to the $T$ parameter. For regions where $m_h \sim 125~\gev$, we find that
$\Tneut < 0.03$.

  %%%%%%%%%%%%%%%%%%%%%%%%%%%%%%%

\begin{comment}
   Hence, generally speaking, with all else considered the same, the discovery prospects of singlino DM are higher than that of bino DM in future direct detection experiments.
\end{comment}

%%%%%%%%%%%%%%%%%%%%%%
\subsection{Bino dark matter}

This parametric scenario is the bino analogue of the previous singlino dark
matter scenario we have discussed. It is also a typical scenario that arises in
the MSSM at large $\tan \beta$. $g_{h\chi\chi}$, in Eq.~(\ref{eq:Couphfull}), will
again be suppressed when $\none$ is mostly bino.~\footnote{As an aside,
any admixture of bino and singlino such that $N^2_{\bino} + N^2_{\sino} \simeq 1$
will also lead to a suppressed $g_{h\chi\chi}$.} The bino fraction of $\none$ is
increased by lowering $M_1$ relative to other mass scales in the neutralino
mass matrix. Again the observed cosmological relic abundance is either through
the mechanisms of resonant annihilation and
co-annihilation~\cite{Griest:1990kh}.

%%%%%%%%%%%%%%%%
\subsubsection{Resonant annihilation region}

Again in the resonant annihilation region, we set $M_1 = 62.5$~GeV, $\mue = 800
$~GeV, $\mu' = 1$~TeV, $\tan\beta = 50$ and $M_A = 4$ TeV. The blue curves in
Fig.~\ref{fig:bisi}(c) depict contours of the LSP-nucleon scattering rates,
$\sigma_{\text{SI}}$ (in units of $\sigma_0 = 10^{-45}~\text{cm}^2$). The remaining
colored contours correspond to the same regions as those in Fig.~\ref{fig:bisi}(a).
The scattering
cross-sections are stronger here than in Fig.~\ref{fig:bisi}(a) because in the singlino-like
scenario there is a partial cancellation among the terms in Eq.~(\ref{eq:Couphfull}), which suppresses
$g_{h\chi\chi}$.
This cancellation arises from an extra minus sign picked up by $N_{\huino}$ for the range of
mass parameters considered.

Similar to Fig.~\ref{fig:bisi}(a), the dark matter-nucleon scattering rates are
seen to decrease as we decouple the Higgsinos by increasing $\mue$. In contrast
to singlino dark matter, $\Tneut \sim 0$ for bino dark matter throughout the
plot in Fig.~\ref{fig:bisi}(c) because both the charged and neutral
Higgsinos are quite degenerate.

%%%%%%%%%%%%%%%%%%%%%
\subsubsection{Co-annihilation region}

Similar to the singlino scenario, we assume that the sfermion mass spectrum is
such the relic density of $\none$ is consistent with cosmological observations.
The blue curves in Fig.~\ref{fig:bisi}(d) depict contours of $\sigma_{\rm SI}$ in
units of $\sigma_0 = 10^{-45}~{\rm cm}^2$.
We vary $M_1$ while fixing $\mu' = 1
$~TeV, and the remaining parameters are the same as in Fig.~\ref{fig:bisi}(b).
The dependence of $\sigma_{\rm SI}$ on $\mu_{\rm eff}$ and $M_1$ is similar to
that of singlino scenario with $\mu' \to M_1$.
Since the Higgsino fraction is
larger in the mostly singlino $\none$ that the mostly bino $\none$,
$\sigma_{\text{SI}}$ is large in Fig.~\ref{fig:bisi}(b) compared to
Fig.~\ref{fig:bisi}(d).
For regions where $m_h \sim 125 \gev$, we find $\Tneut \sim 0$.

 \subsection{The well-tempered bino/singlino/Higgsino}

%%%%%%%%%%%%%%%%%%%%%%%%%%%%%%
\begin{figure}
\begin{center}
\includegraphics[width=0.4\textwidth]{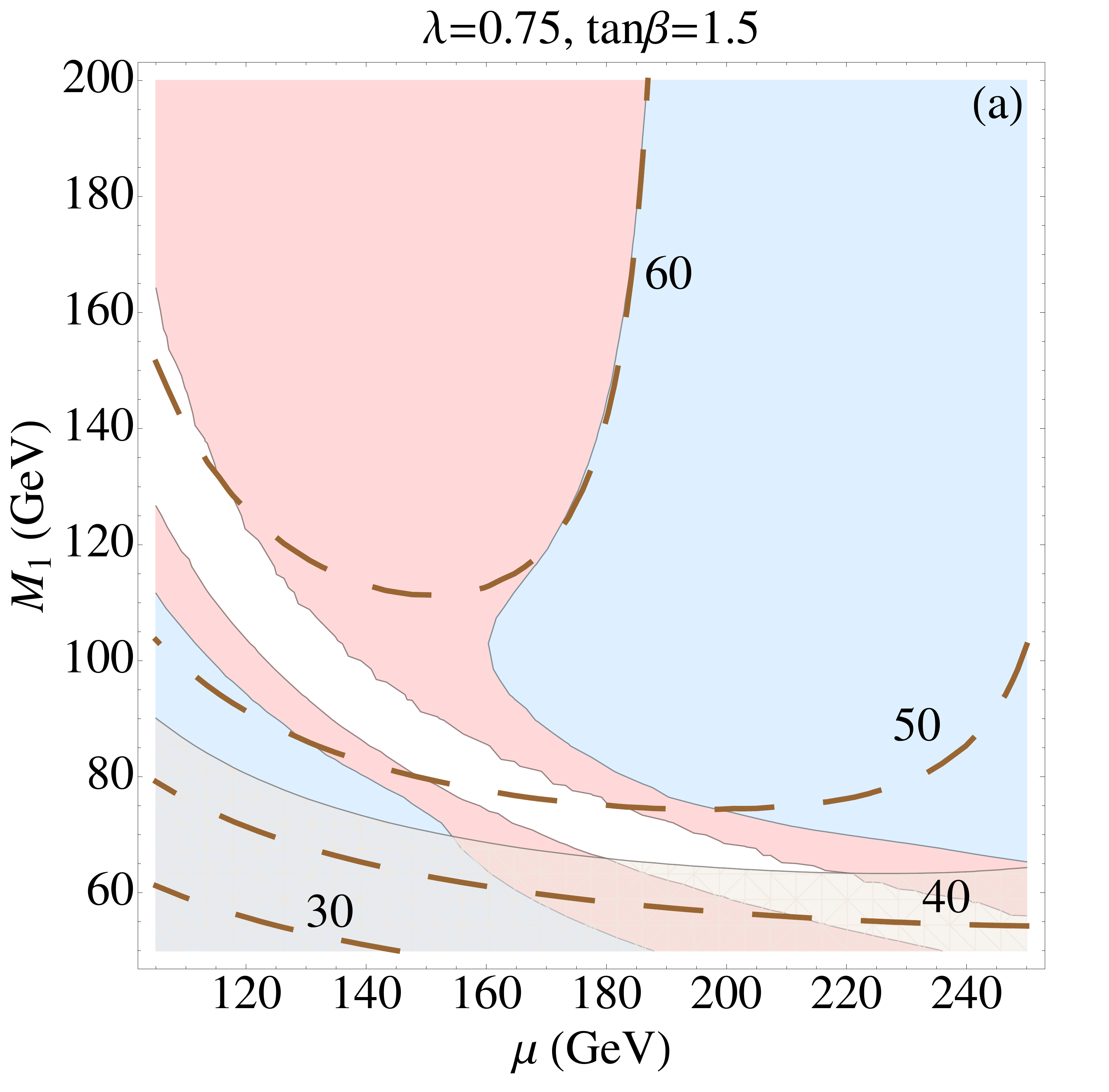}
\quad \quad
\includegraphics[width=0.4\textwidth]{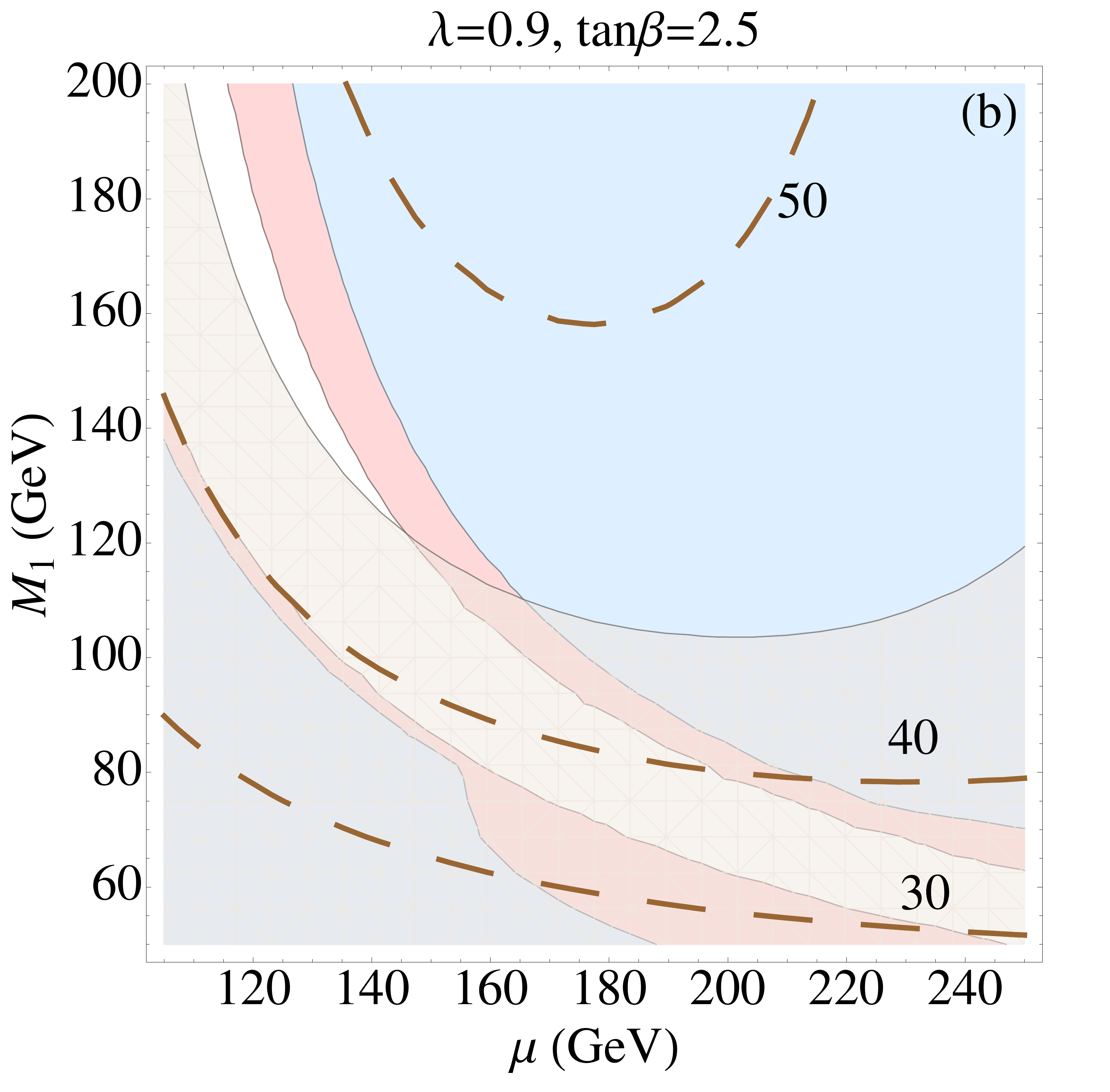}
\caption{\footnotesize The well-tempered scenario at low $\tan \beta$, with $\none$ an admixture of bino, Higgsino and singlino.
In (a), $\lambda = 0.75,~\tan \beta = 1.5$ and in (b), $\lambda = 0.9,~\tan \beta = 2.5$.
The heavy Higgs states are decoupled at $M_A = 5$ TeV.
This choice of parameters fixes $m_h \sim 125$~GeV.
 Regions shaded red are excluded by LUX at 90$\%$~C.L.,
 blue by $h \ra \none~\none$ bounds and gray by $Z \ra \none~\none$ bounds.
These constraints leave a small patch of parameter space that are still viable, the ``blind spots".
The dashed lines are contours of $\Mchione$ in GeV. More details are presented in the text.}
\label{fig:welltemp_bihisi}
\end{center}
\end{figure}
%%%%%%%%%%%%%%%%%%%%%%%%%%%%%%

%  Broadly speaking, at low $\tan\beta$, the Higgs mass can be set at 125 GeV by two different effects %(ignoring corrections from the stop sector). If we define $m^2_{hh} \equiv \lambda^2 v^2 s^2_{2\beta} + M^2_Z % c^2_{2\beta}$, one finds that
  % for $\lambda < 1$, one can decouple the scalar singlet to obtain
 % $m^2_h \simeq m^2_{hh} $, whereas for $\lambda \gsim 1$, one finds $m^2_{hh} > m^2_h$, but $m^2_h = % 125$ GeV can be achieved by mixing with the scalar singlet. We denote the Higgs masses in these two
 % situations as $m_{h,\text{dec}}$ and $m_{h,\text{mix}}$ respectively.

In the limit where $\mu' \ll \mue,~M_1$, precision electroweak contraints can be
evaded by raising $\mue$, thereby decoupling the Higgsinos.
However, raising $\mue$ or $\tan \beta$ suppresses the mass $\none$ as
%%%%%%%%%%%%%%%
\bea
\Mchione \approx \mu' +
\lambda^2 v^2 \mue s_{2\beta}/(\mue^2 + \lambda^2 v^2)
\eea
%%%%%%%%%%%%%%%
for large $M_1$ and $M_2$.
As $\Mchione \leq M_Z/2$ for a large region of
parameters in this scenario, the invisible width of the $Z$ boson is an important
constraint.
Consequently, to find a viable region of parameter space, we require
$\mue \sim \lambda v$ and small $\tan \beta$. In this region, $g_{h\chi\chi}$ is
supppressed when $\none$ is an admixture of $\bino$, $\huino$, and $\sino$ such
that they lead to ``blind spots" in parameter space~\cite{Cheung:2012qy} --
regions that are compatible with current experiment.
%  A key difference between the bino/singlino dark matter discussed above and this % ``well-tempered" scenario is the size of $\mu'$.
%   While we required $\mu' \gsim \lambda v$ in the previous scenario, here we require % $\mu' \ll \lambda v$ in order to achieve large mixing between $\huino$ and $\sino$.
   For illustration, we have consistently set $\mu' = 0$ in this section.
 %  By inspecting the lower $3 \times 3$ block of Eq.~(\ref{eq:massmat_neutralinos}), it % is seen that increasing $\tan\beta$ typically decreases the LSP mass.
 %  For a large enough $\tan\beta$, $\Mchione < M_Z/2$, bringing the limits from the % invisible decay of $Z$ into play, ruling out such regions.

    We illustrate these blind spots in Fig.~\ref{fig:welltemp_bihisi}, which
shows constraints on the LSP in the $M_1 - \mue$ plane.
Fig.~\ref{fig:welltemp_bihisi}(a) corresponds to $\tan\beta = 1.5$ and
Fig.~\ref{fig:welltemp_bihisi}(b) to $\tan\beta=2.5$. To fix $m_h \sim 125$ GeV,
we take $\lambda=0.75$ and $\lambda=0.9$ respectively and decouple the heavy
Higgses with $M_A = 5$ TeV. At these values of $\lambda$, $\tan \beta$ and $M_A$
the Higgs mass is mainly set by the tree-level values as the loop level
corrections are small. The regions shaded red are excluded by LUX at 90\%~C.L.
Regions shaded blue are excluded by the latest limit on the invisible decay of
the Higgs, $\mathcal{B.R.}(h \rightarrow \none\none) < 0.44$~\cite{Chatrchyan:2014tja,Khachatryan:2014jba}.
The gray region is excluded by limits from the invisible decay of the $Z$.
The dashed curves represent contours of $\Mchione$ in GeV. This range of parameters is cosmologically viable with $\Omega_\chi h^2 < 0.12$, where the dominant primordial annihilation of $\none$ is through an $s$-channel $Z$.

    A comparison across the plots informs us that an increase in $\tan\beta$ strengthens the constraints from $Z,h \rightarrow \none\none$, which is due to the decrease in $\Mchione$, as discussed earlier. We also notice that the LUX constraints are consistently stronger than $h \ra \none \none$ bounds. Therefore, the  blind spots (unshaded regions) are determined in this case by limits from LUX and invisible $Z$ decays alone.
 As mentioned in Sec.~\ref{subsec:neut_sector}, larger values of $\lambda$ contribute more to $\Tneut$.
 For Fig.~\ref{fig:welltemp_bihisi}(a) and  Fig.~\ref{fig:welltemp_bihisi}(b), $\Tneut < 0.02$ (completely safe) and $\Tneut < 0.11$ (marginally safe) in the blind spots.
Also, using NMSSMTools 4.5.1 we find that  $\Omega_\chi h^2 \sim 0.01$ in the blind spots.
Therefore, the well-tempered neutralino here can only make up a small fraction of the observed relic density.

%%%%%%%%%%%%%%%%%%%
\subsection{Future prospects}

In this region (large $\tan \beta$ with $\mu' \sim \lambda v \sim \mue$),
the non-standard Higgs scalars are heavy with $M_A$ between $4-8~\tev$.
Therefore the doublet-singlet mixing in the Higgs sector is very small leading
to a ~1$\%$ deviation in the Higgs signal strengths from the SM. Such
deviations are much below the sensitivity of the LHC at present and future
runs, and can only be tested at a future ``Higgs factory". However, the large
$\tan \beta$ scenarios can be probed by future dark matter direct detection
experiments. In particular, the projected reach of the XENON1T
experiment~\cite{Aprile:xenon1T} corresponds to $\sigma_{\rm SI} \approx 10^{-47}-
10^{-46}~{\rm cm}^2$ for dark matter masses between $50$~GeV and $500$~GeV.
Since the DM-nucleon scattering cross-sections for the large $\tan \beta$
scenarios in Fig.~\ref{fig:bisi}  vary from $\sim 10^{-46}-10^{-45}~{\rm cm}^2$,
these regions can be probed at the XENON1T experiment. Unlike the large $\mu',
\tan \beta$ scenarios, XENON1T will
only be able to probe some of the allowed regions of the well-tempered scenario because $g_{h\chi \chi}$ in
can be suppressed.

  %%%%%%%%%%%%%%%%%%%%%%%%%%%%%%%%

\section{Conclusions}
\label{sec:conclusion}

In this article we have investigated the viability of regions of large $\tan
\beta$ in the framework of Fat Higgs/$\lambda$-SUSY models.
In the ``toy'' model we constructed we showed that the singlet cubic term is suppressed while the the tadpole and singlino mass parameter terms were allowed.
Within this framework we showed that there are regions of large $\tan \beta$ that are phenomenologically viable.

In particular, we computed the one-loop effective potential and showed that the $\tan \beta$-independent contributions to the Higgs quartic are crucial in raising the Higgs mass to the observed value of 125~GeV.
We have also shown that non-standard Higgs bosons of the same mass as the stops will give comparable contributions to the Higgs quartic when $\lambda \simeq \sqrt{3} y_t$.
In the region of degenerate non-standard Higgs boson masses the corrections are larger than those estimated in Ref.~\cite{Ellwanger:2005fh,Ellwanger:2009dp}.
This discrepancy is purely due to the assumptions made in Ref.~\cite{Ellwanger:2005fh,Ellwanger:2009dp} that lead to a split spectrum of heavy CP-even and CP-odd scalars.

Furthermore, we pointed out that contributions of the neutralinos and charginos to electroweak precison observables are small even for large $\tan \beta$ when $\mu_{\rm eff} \simeq 500$~GeV and $\mu' \gsim 100$~GeV.
 Such large values of $\mu_{\rm eff}$ make this region of Fat Higgs/$\lambda$-SUSY parameter space slightly more unnatural than the low $\tan \beta$ region considered in Ref.~\cite{Barbieri:2006bg,Cao:2008un,Franceschini:2010qz,Hall:2011aa,Kanemura:2012uy,Perelstein:2012qg,Kyae:2012df,Gherghetta:2012gb,Barbieri:2013hxa,Farina:2013fsa,Gherghetta:2014xea,Zheng:2014loa,Cao:2014kya}. Additionally, this scenario corresponds to the decoupling limit where the mixing between the heavy Higgs states and the SM-like Higgs is suppressed.
 Consequently, SM-like Higgs decay properties are with 1\% of their corresponding Standard Model values.
 Detecting this scenario at the LHC, therefore, will be challenging.

We also found regions of large $\tan \beta$ in Fat Higgs/$\lambda$-SUSY models
that satisfy all the above constraints and provide a viable dark matter
candidate.
For large $\mu'$ and $\tan \beta$ we showed that four possible
viable parametric scenarios exist.
The dark matter in these scenarios could be
either most singlino or bino and, depending on their mass, could generate the
observed relic abundance through resonant annihilation or co-annihilation.
In each of these scenarios, direct detection cross-section can be probed at the
XENON1T experiment.
Another possibility is that of a well-tempered neutralino.
This scenario typically occurs at low values of $\tan \beta$ and $\lambda \lsim 1$,
where the lightest neutralino's Higgsino, bino and singlino fractions are such that its
coupling to the Higgs boson is suppressed.
The XENON1T experiment may not be able to completely probe this scenario.

%%%%%%%%%%%%%%%%%%%%%%%%%%%
\section*{Acknowledgments}

We are grateful to
Kaustubh Agashe,
Spencer Chang,
Graham Kribs 
and Carlos Wagner
for helpful discussions, and
particularly SC and GK for reading our manuscript.
NR thanks Notre Dame University, Michigan State University
and University of Maryland for hospitality, where part of this work was completed.
Our work at University of Oregon was supported in part by the US Department of Energy
under contract number DE-SC0011640.
%%%%%%%%%%%%%%%%%%%%%%%%%%%

%%%%%%%%%%%%%%%%%%%%%%%%%%%%%%%%%%%%
%\section*{Appendix}

\appendix

\section{Decoupling behavior at one-loop level}
\label{apx:decoupling}

We need to use the tadpoles at the one-loop level to solved for the one-loop corrected soft squared mass parameters.  Extending Eq.~(\ref{eq:tadpoles_tree}) to one-loop order leads to the system of three equations,
%%%%%%%%%%%%%%
\bea
 T_i = \frac{\partial V_{\rm Higgs}}{\partial \phi_i} &=& \left. \frac{\partial V^{\rm tree}_{\rm Higgs}}{\partial \phi_i}\right|_{\{v_k\}}   +  \left. \frac{\partial \Delta V}{\partial \phi_i} \right|_{\{v_k\}} = T^{\rm tree}_i + \Delta T_i =  0,  \ \ \ \ \ \ \ \ i=1,2,3.
 \label{eq:tadpoles_loop}
 \eea
 %%%%%%%%%%%%%
We again can try to solve for the soft masses $m^2_{H_u}, m^2_{H_d}$ and $m^2_S$ in terms of the Higgs VEVs. Note that while each $T^{\rm tree}_i$, as given in Eq.~(\ref{eq:tadpoles_tree}), contains only its corresponding soft mass $m^2_{H_i}$, $\Delta T_i$  in general contain all three soft mass terms. Although obtaining the solutions to such a system of equations maybe straightforward,  the computation could become complicated when we expand the full potential around the true electroweak symmetry breaking minimum. We can avoid this difficulty by solving Eq.~(\ref{eq:tadpoles_loop}) iteratively. We first solve for the tree level soft mass squared parameters $\left(m_{H_u^0}^2\right)^0,\left(m_{H_d^0}^2\right)^0,\left(m_S^2\right)^0$ using Eq.~(\ref{eq:tadpoles_tree}) and then substitute them into $\Delta T_i$. This approximation linearizes Eq.~(\ref{eq:tadpoles_loop}) which leads to the one-loop corrected soft mass squared parameters solution
%%%%%%%%%%%%%%%%%%%%%%%
\bea
m_i^2 = \left(m_i^2\right)^0 - \frac{1}{16\pi^2} \sum_{j=D,S} \frac{ M^2_{A,j}}{v_i} \left.\frac{\partial b_j^0}{\partial \phi_i}\right|_{v_i} + ...
\eea
%%%%%%%%%%%%%%%%%%%%%%
where $\left(m_i^2\right)^0$ is the tree-level solutions of Eq.~(\ref{eq:tadpoles_tree}), $b^0_j = b_j \left( (m^2_i)^0 \right)$, $v_i = (v_u,v_d,s) $, $i = (H_u^0,H_d^0,S)$ and $\phi_i =  (H_u^0,H_d^0,S)$. Substituting these solutions into the total potential and expanding it about the electroweak symmetry breaking minimum we observe that corrections to the CP-even Higgs mass matrix takes the form
%%%%%%%%%%%%%%%
\bea
\left(\Delta M_{H^0}^2\right)^{ab} = \frac{1}{16\pi^2} \sum_{i=D,S} \left(\left. \frac{\partial {b^0_{i}}^2}{\partial \phi_a \partial \phi_b}\right|_{\{v_a\}} - \left. \frac{1}{2v_a}\frac{\partial b^0_i}{\partial \phi_a} \right|_{\{v_a\}} \delta_{ab} \right) M_{A,i}^2 + ...,
\label{eq: MA2coeff_post_tadpole_elim}
\eea
%%%%%%%%%%%%%%%
By the symmetries of the model, the only field dependences at quadratic order in $b^0_i$ are $h^2_u, h^2_d, h_u h_d$ and $h^2_s$.
Thus Eq.~(\ref{eq: MA2coeff_post_tadpole_elim}) suggests that the coefficient of $\MAtwosq$ in the self-energy corrections vanishes and that of $\MAonesq$ will be proportional to
%%%%%%%%%%%%%%
\bea
-\frac{v^2}{s_\beta c_\beta}  \left(
 \begin{array}{ccc}
   c^2_\beta  &  - s_\beta c_\beta & 0  \\
 - s_\beta c_\beta  &  s^2_\beta & 0 \\
 0  & 0 & 0
                \end{array} \right).
\eea
%%%%%%%%%%%%%%
When these correction are rotated into the basis defined in Eq.~(\ref{eq:bogdanbasis}) we see that the $(2,2)$ element is the only
non-zero element. Therefore the decoupling is manifest even at the one-loop level.

\section{Effective potential derivation}
\label{apx:effpotderiv}

   In this section we apply the procedure outlined in Sec.~\ref{subsec:higgs_sector} to the computation of one-loop radiative corrections from the Higgs sector. First, we deal with degenerate pseudoscalars, so that all the one-loop corrections come from a single heavy scale. We will call this Case (A). Next, in Case (B), we inspect the effect of splitting the pseudoscalar masses on the one-loop corrections, where they now come from two heavy scales. For simplicity, the soft terms $A_\lambda, A_\kappa, \mu', m_3, m_S'$ are taken to vanish in this case.
\\

(A)~\textbf{Degenerate pseudoscalars}
\\

   From the CP-odd mass matrix in Eq.~(\ref{eq:massmatcpodd}), we impose the necessary and sufficient condition for mass degeneracy in the pseudoscalars given by $\left(M^2_A\right)_{12} = 0, \left(M^2_A\right)_{11} = \left(M^2_A\right)_{22} = M^2_A$, to obtain
  %%%%%%%%%%%%%%%%%%
  \bea
   \mu' &=& A_\lambda, \nonumber \\
  \xi_F &=& (M^2_A s_\beta c_\beta - m^2_3)/\lambda - 2 A_\lambda s, \nonumber \\
   \xi_S &=& -M^2_A s -
  A_\lambda(M^2_A s_\beta c_\beta - m^2_3-\lambda v^2 s_{2\beta})/\lambda
  \eea
  %%%%%%%%%%%%%%%%%%%%%
Respecting this condition, the field-dependent mass matrix for the charged sector is
 %%%%%%%%%%%%%%%%%%
  \bea
{M_{11}^\pm}^2  &=& m_{H_u}^2 + \lambda^2 h_s^2 + \frac{g^2}{4} (h_u^2 - h_d^2) + \frac{g_2^2}{2} h_d^2, \nonumber \\
  {M_{12}^\pm}^2 &=& ( \frac{g_2^2}{2}-\lambda^2)h_u h_d + 2 \lambda A_\lambda (h_s - s)  + M^2_A s_\beta c_\beta, \nonumber \\
  {M_{22}^\pm}^2 &=&m_{H_d}^2 + \lambda^2 h_s^2 - \frac{g^2}{4} (h_u^2 - h_d^2) + \frac{g_2^2}{2} h_u^2,
  \eea
  %%%%%%%%%%%%%%%%%%%%%
for the CP-odd sector it is
 %%%%%%%%%%%%%%%%%%
  \bea
{M_{11}^P}^2  &=& m_{H_u}^2 + \lambda^2 (h_d^2 + h_s^2) + \frac{g^2}{4} (h_u^2 - h_d^2) , \nonumber \\
  {M_{12}^P}^2 &=&  2\lambda A_\lambda (h_s - s) + M^2_A s_\beta c_\beta, \nonumber \\
  {M_{22}^P}^2 &=&  m_{H_d}^2 + \lambda^2 (h_u^2 + h_s^2) - \frac{g^2}{4} (h_u^2 - h_d^2), \nonumber \\
  {M_{13}^P}^2  &=& 0, \nonumber \\
  {M_{23}^P}^2 &=& 0, \nonumber \\
  {M_{33}^P}^2 &=& m_{S}^2 + \lambda^2 (h_u^2 + h_d^2) + A^2_\lambda - {m'_s}^2 , \nonumber \\
  \eea
  %%%%%%%%%%%%%%%%%%%%%
and for the CP-even sector it is
 %%%%%%%%%%%%%%%%%%
  \bea
{M_{11}^S}^2  &=& m_{H_u}^2 + \lambda^2 (h_d^2 + h_s^2) + \frac{g^2}{4} (3 h_u^2 - h_d^2), \nonumber \\
  {M_{12}^S}^2 &=&   (2 \lambda^2 - \frac{g^2}{2}) h_u h_d - 2\lambda A_\lambda (h_s - s) + M^2_A s_\beta c_\beta, \nonumber \\
  {M_{22}^S}^2 &=&  m_{H_d}^2 + \lambda^2 (h_u^2 + h_s^2) - \frac{g^2}{4} (h_u^2 - 3 h_d^2), \nonumber \\
  {M_{13}^S}^2 &=& 2 \lambda^2 (h_u h_s - A_\lambda h_d), \nonumber \\
  {M_{23}^S}^2 &=& 2 \lambda^2 (h_d h_s - A_\lambda h_u) , \nonumber \\
  {M_{33}^S}^2 &=&  m^2_S + \lambda^2 (h_u^2 + h_d^2) + A^2_\lambda +  {m'_s}^2 \nonumber \\
  \label{eq:fielddep_massmat_cpeven}
  \eea
%%%%%%%%%%%%%%%%%%%%%%%%%%%%%%%%

 The eigenvalues of the charged matrix are given by ${M_{1,2}^\pm}^2 = \frac{1}{2}(\text{Tr}_c \mp \sqrt{\text{Tr}_c^2 - 4 \text{Det}_c})$, where $\text{Tr}_c = {M_{11}^\pm}^2 + {M_{22}^\pm}^2$ and $\text{Det}_c = {M_{11}^\pm}^2 {M_{22}^\pm}^2 - {M_{12}^\pm}^2 {M_{21}^\pm}^2$. We only include the contribution from the heavier eigenstate corresponding to ${M_2^\pm}^2$.
Note that when we take the supertrace in the charged higgs sector, we obtain a multiplicative factor of
2 since each charged higgs state comprises of two real physical states. In other words, the supertrace is here taken over the full $4\times4$ squared-mass matrix and not the $2\times2$ version that is usually written down for brevity.

The eigenvalues of the CP-odd matrix are obtained in a straightforward manner, since the upper left $2\times2$ block is decoupled from ${M_{33}^P}^2$. The squared eigenmasses are obtained as $M^2_{1,p} = \frac{1}{2}(\text{Tr}_p - \sqrt{\text{Tr}_p^2 - 4 \text{Det}_p}), M^2_{2,p} = \frac{1}{2}(\text{Tr}_p + \sqrt{\text{Tr}_p^2 - 4 \text{Det}_p})$ and $M^2_{3,p} = M^2_{33}$, where $\text{Tr}_p = {M^P_{11}}^2 + {M^P_{22}}^2$ and $\text{Det}_p = {M^P_{11}}^2 {M^P_{22}}^2 - {M^P_{12}}^2 {M^P_{21}}^2$.

Obtaining the CP-even eigenvalues is non-trivial since we need to deal with a rank 3 matrix. However, we can take advantage of the degeneracy of the CP-odd scalars by
employing the following simplifying trick.

 First, consider the characteristic
equation of the CP-even matrix, written as
%%%%%%%%%%%%%%
\bea
\alpha_3 x^3 + \alpha_2 x^2 + \alpha_1 x + \alpha_0 = 0, \nonumber
\eea
%%%%%%%%%%%%%%%
whose solutions are the field-dependent eigenmasses $M^2_{i,s}$. The coefficients $\alpha_i$, in terms of the matrix elements in Eq.~(\ref{eq:fielddep_massmat_cpeven}), are
%%%%%%%%%%%%%%%%%%%%%%%%%%%%%%%%%%
\begin{eqnarray}
 \alpha_3 &=& 1, \nonumber \\
\alpha_2 &=& -({M_{11}^S}^2 + {M_{22}^S}^2 +  {M_{33}^S}^2), \nonumber  \\
 \alpha_1 &=& {M_{11}^S}^2{M_{22}^S}^2 + {M_{22}^S}^2{M_{33}^S}^2 + {M_{33}^S}^2{M_{11}^S}^2
 - {M_{12}^S}^2{M_{21}^S}^2 - {M_{23}^S}^2{M_{32}^S}^2 - {M_{31}^S}^2{M_{13}^S}^2, \nonumber  \\
 \alpha_0 &=&  -[
                   {M_{11}^S}^2 ({M_{22}^S}^2{M_{33}^S}^2-{M_{23}^S}^2{M_{32}^S}^2)
                  -{M_{12}^S}^2 ({M_{21}^S}^2{M_{33}^S}^2-{M_{23}^S}^2{M_{31}^S}^2) \nonumber \\
                  & & ~~+ {M_{13}^S}^2 ({M_{21}^S}^2{M_{32}^S}^2-{M_{22}^S}^2{M_{31}^S}^2)
                  ]
                  \label{eq:charac_eq_coeffs_gauges}
\end{eqnarray}
%%%%%%%%%%%%%%%%%%%%%%%%%%%%%%%%%
We also know, in terms of the eigenmasses, that
%%%%%%%%%%%%%%%%%%%%
\bea
 \alpha_2 &=& -(M^2_{1,s}+M^2_{2,s}+M^2_{3,s}), \nonumber \\
  \alpha_1 &=& M^2_{1,s}M^2_{2,s} + M^2_{2,s}M^2_{3,s} + M^2_{3,s}M^2_{1,s}
  \label{eq:charac_eq_coeffs_eigens}
\eea
%%%%%%%%%%%%%%%%%%%%

Now the CP-even sector contribution to the effective potential, from Eq.~(\ref{eq:DeltaV_full}), is
%%%%%%%%%%%%%%%%%%%%%%%%%%%%%%%%%
\bea
\Delta V \supset \frac{1}{64 \pi^2} [(M^2_{2,s})^2 + (M^2_{3,s})^2] \log\left(\frac{M^2_A}{M^2_Z}\right).
\label{eq:Veff_cpevencontrib}
\eea
%%%%%%%%%%%%%%%%%%%%%%%%%%%%%%%%%%

The quantity in brackets can be re-written using Eq.~(\ref{eq:charac_eq_coeffs_eigens}) as simply
%%%%%%%%%%%%%%%%%%%%%%%%%%%%%%%%%
\bea
(M^2_{2,s})^2 + (M^2_{3,s})^2  = \alpha^2_2  -  2 \alpha_1 - (M^2_{1,s})^2
\label{eq:term_in_brackets}
\eea
%%%%%%%%%%%%%%%%%%%%%%%%%%%%%%%%%%
The coefficients $\alpha_1$ and $\alpha_2$ may be read off Eq.~(\ref{eq:charac_eq_coeffs_gauges}), while we may still have to determine $M^2_{1,s}$ analytically. This is, however, a simple task if we write $M^2_{1,s}$ as a power series in $M^2_A : $
%%%%%%%%%
\bea
 M^2_{1,s} &=& b_1 + \mathcal{O}\left(\frac{1}{M^2_A}\right) \nonumber \\
\implies (M^2_{1,s})^2 &=& b^2_1 + \mathcal{O}\left(\frac{1}{M^2_A}\right) ,
\label{eq:M1s_expansion}
\eea
%%%%%%%%%
where $b_1$ is at most quadratic in the background fields. Putting Eqs.~\ref{eq:term_in_brackets} and \ref{eq:M1s_expansion} into Eq.~(\ref{eq:Veff_cpevencontrib}), we obtain the one-loop effective potential contribution simply as
%%%%%%%%%%%%
\bea
\Delta V \supset \frac{1}{64 \pi^2} [\alpha^2_2  -  2 \alpha_1 - b^2_1] \log\left(\frac{M^2_A}{M^2_Z}\right),
\eea
%%%%%%%%%%%%

 where we have discarded $\mathcal{O}(1/M^2_A)$ terms that are irrelevant in obtaining the required self-energy corrections.

 After including all the one-loop corrections, the final expressions we obtain for
the CP-even mass matrix are now as follows.

%%%%%%%%%%%%%%%%
\bea
\left(\overline{M}^2_H\right)_{11} = M^2_Z s^2_\beta +   \overline{M}^2_A  c^2_\beta + \Pi_{11} ; &\;\;\;\;&
\left(\overline{M}^2_H\right)_{12} = (2\lambda^2 v^2 - M^2_Z - \overline{M}^2_A) s_\beta c_\beta + \Pi_{12}; \nonumber \\
\left(\overline{M}^2_H\right)_{22} = M^2_Z c^2_\beta +  \overline{M}^2_A+ \Pi_{22}; &\;\;\;\;\;&
\left(\overline{M}^2_H\right)_{13} =  2\lambda v \mue s_\beta + \Pi_{13};  \nonumber \\
\left(\overline{M}^2_H\right)_{23} = 2\lambda v \mue c_\beta + \Pi_{23}; &\;\;\;\;\;&
\left(\overline{M}^2_H\right)_{33} = \overline{M}^2_A + \Pi_{33},  \nonumber
\eea
%%%%%%%%%%%%%%%%
where
%%%%%%%%%%%%%%%%
\bea
\overline{M}^2_A &=& M^2_A \left(1 + \frac{\lambda^2}{8\pi^2}   \log\left(\frac{M^2_A}{M^2_Z}\right) \right),  \eea
%%%%%%%%%%%%%%%%
and
%%%%%%%%%%%%%%%%
\bea
\Pi_{11} &=& \frac{v^2}{256\pi^2} [- 32 \lambda^4 s_\beta^2 (2 c_{2\beta}-s_{2\beta}^2)
 +2  \lambda^2  g^2(3 c_{2\beta}-1) (3 s_{2\beta}^2+2) \nonumber \\
& & + g^4  (4 c_W^4+4 c_W^2-7 s_{2\beta}^2-1-c_{2\beta} (4 c_W^4-4 c_W^2+5 s_{2\beta}^2+3 )) \nonumber \\
 & & + 64\lambda^2 \frac{\Al \mue}{v^2} \cot\beta ] \log\left(\frac{M^2_A}{M^2_Z}\right), \nonumber \\
\Pi_{12} &=&  \frac{v^2}{256\pi^2} [- 32  \lambda^4  (s_{2\beta}^2-2) - 2 \lambda^2 g^2 s_{2\beta} (8 c_W^2-15 s_{2\beta}^2+14 )\nonumber \\
 & & + g^4  s_{2\beta}(4 c_W^4+4 c_W^2-7 s_{2\beta}^2+3 ) \nonumber \\
 & &  - 64\lambda^2 \frac{\Al \mue}{v^2} ] \log\left(\frac{M^2_A}{M^2_Z}\right), \nonumber \\
  \nonumber \Pi_{22} &=&  \frac{v^2}{256\pi^2} [ 32 \lambda^4 c_\beta^2  (2 c_{2\beta}+s_{2\beta}^2) -2  \lambda^2 g^2  (3 c_{2\beta}+1) (3 s_{2\beta}^2+2 ) \nonumber \\
& & + g^4  (4 c_W^4-4 c_W^2+7 s_{2\beta}^2+1 + c_{2\beta}  (4 c_W^4-4 c_W^2+5 s_{2\beta}^2+3 ) ) \nonumber \\
& & + 64\lambda^2 \frac{\Al \mue}{v^2} \tan\beta]
  \log\left(\frac{M^2_A}{M^2_Z}\right), \nonumber \\
 \Pi_{13} &=& \frac{v \mue}{\mue} [ 12 \lambda^3  s_\beta^3 + \lambda g^2 s_\beta (3 c_{2\beta}+2 c_W^2+1 )  \nonumber \\
& & + \frac{\lambda v \Al \cos \beta }{32\pi^2} \left( - \lambda^2 (13 + 3 c_{4\beta}) + \frac{g^2}{2} (5 + 4c^2_W - 6 c_{2\beta} + 3 c_{4\beta}) \right)] \log\left(\frac{M^2_A}{M^2_Z}\right),  \nonumber  \\
  \nonumber \Pi_{23} &=& \frac{v \mue}{\mue} [ 12 \lambda^3 c_\beta^3 + \lambda g^2 c_\beta (-3 c_{2\beta}+2 c_W^2+1 ) ) \nonumber \\
  & & + \frac{\lambda v \Al \sin \beta }{32\pi^2} \left( - \lambda^2 (13 + 3 c_{4\beta}) + \frac{g^2}{2} (5 + 4c^2_W + 6 c_{2\beta} + 3 c_{4\beta}) \right)] \log\left(\frac{M^2_A}{M^2_Z}\right),
 \nonumber  \\
 \Pi_{33} &=& \{ \frac{4 \lambda^2 \mue^2}{16\pi^2} + \frac{\lambda \Al}{128\pi^2} [\lambda (16 \Al (4+c_{4\beta}) \nonumber \\
  & & + \lambda (64s^2 + 29 v^2)s_{2\beta} + \lambda v^2s_{6\beta} )  + g^2 v^2  s_{2\beta} (3+4 c^2_W + c_{4\beta})] \}   \log\left(\frac{M^2_A}{M^2_Z}\right)
  \end{eqnarray}
%%%%%%%%%%%%%%%

If we set all NMSSM-specific parameters to zero in the above, we recover the MSSM limit
presented in \cite{Haber:1993an,Gladyshev:1994iw,Pierce:1996zz,Dobado:2002jz}.
The soft term $\Al$ decouples at one-loop order and does not contribute to the SM Higgs quartic coupling, a property best seen in the basis
 %$H_1 = H_u s_\beta + H_d c_\beta,~ H_2 = H_u c_\beta - H_d s_\beta,~S = S$, that we discussed in Sec.~\ref{subsec:higgs_sector}.
 of Eq.~(\ref{eq:bogdanbasis}).
 % We must
 % rotate the self-energy corrections $\Pi_{ij}$ ($i,j = 1,2,3$) into this basis: $\Pi_{\beta,ij} = (R_\beta)_{ai} \Pi_{ab}(R_\beta)_{jb}$, where $R_\beta$ is as defined in Eq.~(\ref{eq:Bogdanrotation}).
 The SM Higgs boson mass is then identified as
%%%%%%%%%%%%%%%%%%%%%%%%%
  \bea
  \overline{M}^2_{hh} &=&  \lambda^2 v^2 s^2_{2\beta} +  M^2_Z c^2_{2\beta} + \Pi_{hh}, \nonumber \\
  \Pi_{hh} &=& \frac{v^2}{512\pi^2} [4\lambda^4(31+4c_{4\beta}-3c_{8\beta}) + 4\lambda^2g^2(-9-4c^2_W+(4c^2_W-2)c_{4\beta}+3c_{8\beta}) \nonumber \\
  & & -g^4(-11+8c^2_W-16c^4_W+8c^2_W c_{4\beta}+3c_{8\beta})] \log \left( \frac{M^2_A}{M^2_Z} \right)
 \eea
%%%%%%%%%%%%%%%%%%%%%%

$\Al$ is absent in the expression above, confirming its decoupling behavior at the one-loop level. Moreover, if we neglect
the electroweak strength corrections, in the limit of large $\tan \beta$  we get
%%%%%%%%%%%%%%%%%%%%
\bea
 \lim_{\tan \beta \gg 1} \Pi_{hh} = \frac{\lambda^4 v^2}{4 \pi^2}\log \left( \frac{M^2_A}{M^2_Z} \right),
\eea
%%%%%%%%%%%%%%%%%%%%
in agreement with our heuristic estimate in Eq.~(\ref{eq:loopestimate_higgses}).
\\

(B)~\textbf{Non-degenerate pseudoscalars: a simple case.}
\\

We now show the effect of a split pseudoscalar spectrum on the radiative corrections. For simplicity,
we assume the parameters $\Al, \mu', m_3, m'_S$ vanish. We also neglect $g$-dependent terms in the one-loop piece,
since the largest contributions to the SM Higgs quartic in our model arise from the $\lambda$-dependent terms. With
these simplifications, the field-dependent squared mass matrices for the charged,
CP-odd and CP-even sectors are respectively given by
%%%%%%%%%%%%%%%%%%%%%%%%%%%%%%%
 \bea
{M_{11}^\pm}^2 = m_{H_u}^2 + \lambda^2 h_s^2,~~~~
{M_{12}^\pm}^2 = \lambda^2 h_u h_d + \MAonesq s_\beta c_\beta,~~~~
{M_{22}^\pm}^2 = m_{H_d}^2 + \lambda^2 h_s^2;
  \eea
  %%%%%%%%%%%%%%%%%%%%%
 %%%%%%%%%%%%%%%%%%
  \bea
{M_{11}^P}^2  = m_{H_u}^2 + \lambda^2 (h_d^2 + h_s^2),~~~~& &
  {M_{12}^P}^2 =  \MAonesq s_\beta c_\beta, \nonumber \\
  {M_{22}^P}^2 =  m_{H_d}^2 + \lambda^2 (h_u^2 + h_s^2),~~~~& &
  {M_{13}^P}^2 = 0, \nonumber \\
  {M_{23}^P}^2 = 0,~~~~& &
  {M_{33}^P}^2 = m_{S}^2 + \lambda^2 (h_u^2 + h_d^2); \nonumber \\
  \eea
  %%%%%%%%%%%%%%%%%%%%%
%%%%%%%%%%%%%%%%%%%
  \bea
{M_{11}^S}^2  = m_{H_u}^2 + \lambda^2 (h_d^2 + h_s^2),~~~~& &
  {M_{12}^S}^2 =  2 \lambda^2 h_u h_d - \MAonesq s_\beta c_\beta, \nonumber \\
  {M_{22}^S}^2 =  m_{H_d}^2 + \lambda^2 (h_u^2 + h_s^2),~~~~& &
  {M_{13}^S}^2 = 2 \lambda^2 h_u h_s, \nonumber \\
  {M_{23}^S}^2 = 2 \lambda^2 h_d h_s,~~~~& &
  {M_{33}^S}^2 =  m^2_S + \lambda^2 (h_u^2 + h_d^2); \nonumber \\
    \eea
%%%%%%%%%%%%%%%%%%%%%%%%%%%%%%%%%

Obtaining the eigenvalues of the charged and CP-odd systems is straightforward again, as we found in Case (A).
To obtain the eigenvalues of the CP-even matrix, we solve for the roots of its characteristic
equation (a cubic polynomial) as a power series in $\MAonesq$ and $\MAtwosq$.

After collecting the one-loop contributions from all three sectors and summing over them, we obtain the CP-even mass matrix as
%%%%%%%%%%%%%%%%
\bea
\left(\overline{M}^2_H\right)_{11} = M^2_Z s^2_\beta +   \MAonesqbar  c^2_\beta + \Pi_{11} ; &\;\;\;\;&
\left(\overline{M}^2_H\right)_{12} = (2\lambda^2 v^2 - M^2_Z - \MAonesqbar) s_\beta c_\beta + \Pi_{12}; \nonumber \\
\left(\overline{M}^2_H\right)_{22} = M^2_Z c^2_\beta +  \MAonesqbar+ \Pi_{22}; &\;\;\;\;\;&
\left(\overline{M}^2_H\right)_{13} =  2\lambda v \mue s_\beta + \Pi_{13};  \nonumber \\
\left(\overline{M}^2_H\right)_{23} = 2\lambda v \mue c_\beta + \Pi_{23}; &\;\;\;\;\;&
\left(\overline{M}^2_H\right)_{33} = \MAtwosqbar + \Pi_{33}  \nonumber
\eea
%%%%%%%%%%%%%%%%
where
%%%%%%%%%%%%%%%%
\bea
\MAonesqbar &=& \MAonesq \left(1 + \frac{\lambda^2}{8\pi^2}   \log\left(\frac{\MAonesq}{M^2_Z}\right) + \frac{\lambda^2}{8\pi^2} \frac{\mue^2}{M^2_{A_2}-M^2_{A_1}} \log\left(\frac{\MAtwosq}{\MAonesq} \right)\right),  \nonumber \\
\MAtwosqbar &=& \MAtwosq
\eea
%%%%%%%%%%%%%%%%
and
%%%%%%%%%%%%%%%%
\bea
\nonumber \Pi_{11} &=& \frac{\lambda^4 v^2}{16\pi^2} s^2_\beta \left[ - (4 c_{2\beta}+c_{4\beta}+1) \log\left(\frac{\MAonesq}{M^2_Z}\right) +  2 \log\left(\frac{\MAtwosq}{M^2_Z}\right)  \right], \nonumber \\
\nonumber \Pi_{12} &=&  \frac{2\lambda^4 v^2}{16\pi^2} s_{\beta} c_{\beta}  (2+c_{4\beta}) \log\left(\frac{\MAonesq}{M^2_Z}\right), \nonumber \\
  \nonumber \Pi_{22} &=& \frac{\lambda^4 v^2}{16\pi^2} c^2_\beta \left[ -  (-4 c_{2\beta}+c_{4\beta}+1) \log\left(\frac{\MAonesq}{M^2_Z}\right)
  + 2 \log\left(\frac{\MAtwosq}{M^2_Z}\right)  \right], \nonumber \\
    \nonumber \Pi_{13} &=& \frac{\lambda^3 v \mue}{16\pi^2} s_{\beta} \left[ -(1+3 c_{2\beta}) \log\left(\frac{\MAonesq}{M^2_Z}\right)
  + 4  \log\left(\frac{\MAtwosq}{M^2_Z}\right) \right],
  \\
  \nonumber \Pi_{23} &=& \frac{\lambda^3 v \mue}{16\pi^2} c_\beta \left[  -(1-3 c_{2\beta}) \log\left(\frac{\MAonesq}{M^2_Z}\right)
  +  4 \log\left(\frac{\MAtwosq}{M^2_Z}\right) \right],
   \\
  \nonumber \Pi_{33} &=& \frac{4 \lambda^2 \mue^2}{16\pi^2}  \log\left(\frac{\MAonesq}{M^2_Z}\right).
    \\
\end{eqnarray}
%%%%%%%%%%%%%%%

We make the following observations concerning the above expressions. First, notice that in the limit $M_{A_D} = M_{A_S}$, they
 are consistent with the results in Case (A) with $g, \Al \ra 0$.
Second, we observe that corrections from the heavy doublet Higgses are $\beta$-dependent
and those from the heavy singlet Higgses are not, as reflected in the co-efficients of $\log(\MAonesq/M^2_Z)$ and
$\log(\MAtwosq/M^2_Z)$ respectively. Third, there is a marked difference in contributions from the scales
$M_{A_D}$ and $M_{A_S}$ to the SM Higgs quartic, which can be understood in the basis
of Eq.~(\ref{eq:bogdanbasis}).
 Rotating $\Pi_{ij}$ into this basis, the correction to the SM Higgs boson mass is identified as
%%%%%%%%%%%%%%%%%%%%%%%%%%%%%%%%%
\bea
\Pi_{hh} = \frac{\lambda^4 v^2 s_\beta}{16\pi^2} \left[ \left(c^2_\beta(2+c_{4\beta}) - s^2_\beta(1+c_{4\beta}+4c_{2\beta})\right) \log\left(\frac{\MAonesq}{M^2_Z}\right) + 2s^2_\beta \log\left(\frac{\MAtwosq}{M^2_Z}\right)  \right]. \nonumber
\eea
%%%%%%%%%%%%%%%%%%%%%%%%%%%%%%%%%
The difference in the co-efficients of the logarithms are greatest at $\tan\beta \sim 1$, and smallest at $\tan\beta \gg 1$. In the latter limit, we obtain
%%%%%%%%%%%%%%%%%%%%%%%%%%%%%%%%%
\bea
\lim_{\tan \beta \gg 1} \Pi_{hh} = \frac{\lambda^4 v^2}{16\pi^2} \left[ 2 \log\left(\frac{\MAonesq}{M^2_Z}\right) + 2 \log\left(\frac{\MAtwosq}{M^2_Z}\right)  \right], \nonumber
\eea
%%%%%%%%%%%%%%%%%%%%%%%%%%%%%%%%%
which is consistent with our qualitative estimate in Eq.~(\ref{eq:loopestimate_differentscales}).

%%%%%%%%%%%%%%%%%%%%%%%%%%%%%%%%%%%

%%%%%%%%%%%%%%%%%%%%%%%%%%%%%%%%

\end{document}